\documentclass[usenatbib]{mnras}

\usepackage{newtxtext,newtxmath}
 
\usepackage[T1]{fontenc}
\usepackage{ae,aecompl}

\usepackage{graphicx}	% Including figure files
\usepackage{amsmath}	% Advanced maths commands
\usepackage{amssymb}	% Extra maths symbols
\usepackage{siunitx}
\usepackage[version=4]{mhchem}
\usepackage[autostyle=true]{csquotes}
\usepackage{subcaption}
\usepackage{xcolor}
\usepackage{footnote}
\definecolor{darkorange}{RGB}{179,98,0}

\title[]{Sulfur Chemistry in the Atmospheres of Warm and Hot Jupiters}

\author[R. Hobbs et al.]{
Richard Hobbs,$^{1}$\thanks{E-mail: rh567@cam.ac.uk}
Paul B. Rimmer$^{2,3,4}$
Oliver Shorttle,$^{1,2}$
and Nikku Madhusudhan$^{1}$
\\
$^{1}$Institute of Astronomy, University of Cambridge, Cambridge, CB3 0HA, UK\\
$^{2}$Cambridge Earth Sciences, University of Cambridge, Cambridge CB2 3EQ, UK\\
$^{3}$MRC Laboratory of Molecular Biology, Cambridge, CB2 OQH, UK\\
$^{4}$Cavendish Astrophysics, University of Cambridge, Cambridge, CB3 0HE, UK\\
}

\date{Accepted XXX. Received YYY; in original form ZZZ}

\pubyear{2020}

\begin{document}
\label{firstpage}
\pagerange{\pageref{firstpage}--\pageref{lastpage}}
\maketitle

\begin{abstract}
We present and validate a new network of atmospheric thermo-chemical and photo-chemical sulfur reactions. We use a 1-D chemical kinetics model to investigate these reactions as part of a broader HCNO chemical network in a series of hot and warm Jupiters. We find that temperatures approaching $1400\,\mathrm{K}$ are favourable for the production of \ce{H2S} and \ce{HS} around $\mathrm{10^{-3}\,bar}$ at mixing ratios of around $10^{-5}$, an atmospheric level where detection by transit spectroscopy may be possible. At $\mathrm{10^{-3}\,bar}$ and at lower temperatures, down to $1000\,\mathrm{K}$, mixing ratios of \ce{S2} can be up to $10^{-5}$, at the expense of \ce{H2S} and \ce{HS}, which are depleted down to a mixing ratio of $10^{-7}$. We also investigate how the inclusion of sulfur can manifest in an atmosphere indirectly, by its effect on the abundance of non-sulfur-bearing species. We find that in a model of the atmosphere of HD 209458 b, the inclusion of sulfur can lower the abundance of \ce{NH3}, \ce{CH4} and \ce{HCN} by up to two orders of magnitude around $\mathrm{10^{-3}\,bar}$. In the atmosphere of the warm Jupiter 51 Eri b, we additionally find the inclusion of sulphur depletes the peak abundance of \ce{CO2} by a factor of five, qualitatively consistent with prior models. We note that many of the reactions used in the network have poorly determined rate constants, especially at higher temperatures. To obtain an accurate idea of the impact of sulfur chemistry in hot and warm Jupiter atmospheres, experimental measurements of these reaction rates must take place.

\end{abstract}


%%%%%%%%%%%%%%%%%%%%%%%%%%%%%%%%%%%%%%%%%%%%%%%%%%

%%%%%%%%%%%%%%%%% BODY OF PAPER %%%%%%%%%%%%%%%%%%

\section{Introduction}

%----------------------------------------------------------------------------------------

% Define some commands to keep the formatting separated from the content 
\newcommand{\keyword}[1]{\textbf{#1}}
\newcommand{\tabhead}[1]{\textbf{#1}}
\newcommand{\code}[1]{\texttt{#1}}
\newcommand{\file}[1]{\texttt{\bfseries#1}}
\newcommand{\option}[1]{\texttt{\itshape#1}}

%----------------------------------------------------------------------------------------
Sulfur chemistry is known to play an important role in the atmospheric chemistry of planets in our solar system. In Earth's past, it may have lead to the first Snowball Earth event (\citealt{Macdonald2017}), and sulfur isotope ratios can be used as a tracer of Earth's Great Oxidation Event (\citealt{Hodgskiss2019}). Sulfur photochemistry is thought to influence the production of hazes and clouds in the upper atmospheres of solar system planets such as Earth (\citealt{Malin1997}), Venus (\citealt{Zhang2012,Titov2018}), Jupiter (\citealt{Moses1995}) and the moon Io (\citealt{Irwin1999}). Similar hazes, expected to appear in the atmospheres of exoplanets (\citealt{He2020}), would greatly affect their observed spectra. This would limit our ability to examine their atmosphere, and make assessments of their habitability challenging (\citealt{Gao2017}).

Sulfur has also recently been of interest in the field of Jupiter-like exoplanets. A tentative detection of the mercapto radical, \ce{HS}, in the atmosphere of the hot Jupiter WASP-121 b has been made (\citealt{Evans2018}). However, it is still uncertain whether \ce{HS} or another molecule is responsible for the absorption feature seen. The warm Jupiter 51 Eri b has also been an excellent example of the importance of sulfur chemistry. Two groups have produced models of 51 Eri b's atmosphere, one study did not include sulfur chemistry (\citealt{Moses2016}) and the other did include sulfur chemistry (\citealt{Zahnle2016}). Significant differences were found between the two models, in particular in the modelled abundance of \ce{CO2}, but it was unknown whether this was due to sulfur chemistry or other differences in the models. Thus, we believe that including sulfur chemistry in models of exoplanets, both terrestrial and gas-giant, would be valuable in obtaining more accurate pictures of their atmospheric composition.

There are several works that include sulfur chemistry for exoplanet atmospheres. We highlight a subset of these relevant for the atmospheric chemistry of gas giants. \cite{Hu2013} investigate the photochemistry of \ce{H2S} and \ce{SO2} in terrestrial planet and conclude that their direct detection is unlikely due to rapid photochemical conversion to elemental sulfur and sulfuric acid. \cite{Visscher2006} examine thermo-chemical sulfur chemistry for the hot, deep atmospheres of sub-stellar objects and find \ce{H2S} to be the dominant sulfur bearing gas. However, due to the high temperatures and pressures of the atmospheres they are studying, disequilibrium effects such as diffusion or photochemistry are not considered. \cite{Zahnle2009} investigate whether sulfur photochemistry could explain the hot stratospheres of hot Jupiters, and find that UV absorption of \ce{HS} and \ce{S2} may provide an explanation. \cite{Zahnle2016} also investigate whether sulfur could form photochemical hazes in the atmosphere of 51 Eri b, and under what conditions they could be formed. They find that sulfur clouds should appear on many planets where UV irradiation is weak. \cite{Wang2017} create synthetic spectra of their hot Jupiter atmospheric models to determine whether \ce{H2S} and \ce{PH3} could be detected by JWST, and find that \ce{H2S} features should be observable for $\mathrm{T_{eq}} > 1500\mathrm{K}$. \cite{Gao2017} examine the effects of sulfur hazes on the reflected light from giant planets and find that the hazes would mask the presence of \ce{CH4} and \ce{H2O}.

We need models to be able to investigate the effects of sulfur on exoplanet atmospheres. One way of modelling the atmosphere of planets is through the use of chemical kinetics models. These exist in the form of 1-D (e.g., \citealt{Moses2011, Venot2012, Rimmer2016, Tsai2017}), 2-D (e.g., \citealt{Agundez2014}) and 3-D (e.g. \citealt{Drummond2018}) models. The 1-D models take a network of possible chemical reactions within an atmosphere and a temperature profile of the atmosphere to produce abundance profiles of chemical species. These models are frequently combined with prescriptions for diffusion and include a UV flux applied to the top of the model, to allow a full disequilibrium model to be created. Atmospheric models are obtained by computing these models until they reach a converged steady-state condition.

Throughout the last few decades, chemical models of exoplanetary atmospheres have grown more detailed and complex. What had begun as equilibrium models containing limited molecular species made of only a few elements in small chemical networks (e.g., \citealt{Lodders2002,Liang2003,Zahnle2009}) have evolved into large photo-kinetic models containing hundreds of molecular species made of many elements in chemical networks with thousands of reactions (\citealt{Moses2011,Venot2012,Rimmer2016}). The focus of this work is to continue this evolution. We do so by introducing a new element, sulfur, to the model and to the network described in our previous paper (\citealt{Hobbs2019}). In this way we can produce a tested and validated network of sulfur chemistry. In this work we apply the network to hot and warm Jupiters. By doing so, we can investigate the importance of sulfur in the atmospheres of exoplanets in both the context of sulfur itself and the way it impacts carbon and oxygen chemistry. It was found in the work of \cite{Zahnle2016} that sulfur photo-chemistry in the atmosphere of the warm Jupiter 51 Eri b was the source of a large number of radicals that went on to catalyse other chemistry. This was a possible explanation for the difference in the \ce{CO2} abundance found for 51 Eri b between the sulfur-free models of \cite{Moses2016} and the sulfurous models of \cite{Zahnle2016}. We revisit 51 Eri b with our own model, as well as the hot Jupiter HD 209458b, to determine whether we can identify what role sulfur plays in the chemistry of both warm and hot Jupiters.

We begin with a brief overview of the model we're using. In Section \ref{Section3} we validate our network for pure thermo-chemical equilibrium and compare our model against the sulfur model of \cite{Wang2017} and \cite{Zahnle2016}. Section \ref{Section4} contains our analysis of the sulfur chemistry occurring in the atmospheres of hot Jupiters, and shows which pathways in our network lead to such chemistry occurring. In Section \ref{Section5}, we examine how sulfur can affect the chemistry of other, non-sulfur species in both warm and hot Jupiters. We discuss our findings and review what we have discovered in Section \ref{Section6}.

\section{Model Details}
\label{Section2}

To investigate sulfur chemistry in the atmospheres of hot Jupiters, we chose to use \textsc{Levi}, a one-dimensional photo-kinetics code, originally described in \cite{Hobbs2019}. This code was previously developed to model carbon, oxygen and nitrogen chemistry in hot Jupiter atmospheres. It was validated against several other photochemical models of hot Jupiters.

In this work, \textsc{Levi}, is being used to model the atmospheres of Jupiter-like planets. It does so via calculations of:

\begin{itemize}
    \item The interactions between chemical species
    \item The effects of vertical mixing due to eddy-diffusion molecular diffusion and thermal diffusion
    \item Photochemical dissociation due to an incoming UV flux.
\end{itemize}

And uses the input parameters of:
\begin{itemize}
    \item The pressure-temperature (P-T) profile of the atmosphere
\item The eddy-diffusion ($K_{zz}$) profile of the atmosphere
\item The profiles of the UV stellar spectrum
\item The metallicity of the atmosphere
\item The gravity of the planet.
\end{itemize}

 It uses the assumptions of:
 
 \begin{itemize}
\item hydrostatic equilibrium
\item The atmosphere being an ideal gas
\item The atmosphere is small compared to the planet, such that gravity is constant throughout the atmospheric range being modelled.  \end{itemize}

By combining all of these factors, abundance profiles of the atmosphere are computed.

As is typical for codes of this type, \textsc{Levi} finds steady-state solutions for species in the atmosphere by solving the coupled one-dimensional continuity equation:

\begin{equation}
\frac{\partial n_{i}}{\partial t} = \mathcal{P}_{i} - \mathcal{L}_{i} - \frac{\partial \Phi_{i}}{\partial z},
\label{eq:continuity}
\end{equation}
where $n_{i}$ (\si{\per\metre\cubed}) is the number density of species $i$, with $i = 1,...,N$, with $N$ being the total number of species. $\mathcal{P}_{i}$ (\si{\per\metre\cubed\per\second}) and $\mathcal{L}_{i}$ (\si{\per\metre\cubed\per\second}) are the production and loss rates of the species $i$. $\partial t$ (\si{\second}) and $\partial z$ (\si{\metre}) are the infinitesimal time step and altitude step respectively. $\Phi_{i}$ (\si{\per\metre\squared\per\second}) is the upward vertical flux of the species, given by,
\begin{equation}
\Phi_{i} = -(K_{zz}+D_i)n_{t}\frac{\partial X_{i}}{\partial z} + D_i n_i\left(\frac{1}{H_0} - \frac{1}{H_i} - \frac{\alpha_{T,i}}{T}\frac{dT}{dz}\right),
\label{eq:diffusion}
\end{equation}
where $X_{i}$ and $n_{t}$ (\si{\per\metre\cubed}) are the mixing ratio and total number density of molecules such that $n_{i} = X_{i}n_{t}$. The eddy-diffusion coefficient, $K_{zz}$ (\si{\metre\squared\per\second}), approximates the rate of vertical transport and $D_i$ (\si{\metre\squared\per\second}) is the molecular diffusion coefficient of species $i$. $H_0$ (\si{\metre}) is the mean scale height, $H_i$ (\si{\metre}) is the molecular scale height, $T$ (K) is the temperature, and $\alpha_{T,i}$ is the thermal diffusion factor. For the full explanation of how we determine each of these parameters, and solve the equations, refer to \cite{Hobbs2019}.

Previously, we used a subset of the Stand2019 network. This was the STAND2015 network first developed in \cite{Rimmer2016} plus the additional reactions from \cite{Rimmer2019}. We limited the network to only reactions with neutral species containing hydrogen, carbon, nitrogen and oxygen, and used it in conjunction with \textsc{Levi} to model hot Jupiter atmospheres. As part of this work we have developed a new addition to this network, comprised of an additional 185 reactions and 30 species that include the element sulfur. Our sulfur species include molecules that contain up to 2 \ce{H}, 3 \ce{C}, 3 \ce{O} and 2 \ce{S}, except for the allotropes of sulfur that include up to \ce{S8}. These species are tabulated in Table \ref{tab:names}. The reactions we have added were mainly drawn from the NIST chemical kinetics database \footnote{https://kinetics.nist.gov/kinetics/}, with the full tabulation of both reactions and sources available in the table of Appendix \ref{The sulfur network}. 

\begin{table}
\centering
\caption{Chemical names of the sulfur species used in the network}
\label{tab:names}
\begin{tabular}{cc}
\hline
Species & Chemical Name\\
\hline
\ce{S}     & Sulfur                \\
\ce{S2}    & Disulfur              \\
\ce{S3}    & Trisulfur             \\
\ce{S4}    & Tetrasulfur           \\
\ce{S5}    & Pentasulfur           \\
\ce{S6}    & Hexasulfur            \\
\ce{S7}    & Heptasulfur           \\
\ce{S8}    & Octasulfur            \\
\ce{HS}    & Mercapto Radical      \\
\ce{H2S}   & Hydrogen Sulfide      \\
\ce{H2S2}  & Hydrogen Disulfide    \\
\ce{CS}    & Carbon Sulfide        \\
\ce{CS2}   & Carbon Disulfide      \\
\ce{CS2OH} & Thioxomethyl Radical  \\ 
\ce{HCS}   & Thioformyl Radical    \\
\ce{OCS}   & Carbonyl Sulfide      \\
\ce{H2CS}  & Thioformaldehyde      \\
\ce{H2C3S} & Tricarbon Monosulfide \\ 
\ce{CH3SH} & Mercaptomethane       \\
\ce{HSNO}  & Thionylimide          \\ 
\ce{SO}    & Sulfur Monoxide       \\
\ce{SO2}   & Sulfur Dioxide        \\
\ce{SO3}   & Sulfur Trioxide       \\
\ce{S2O}   & Disulfur Monoxide     \\
\ce{HSO}   & Sulfenate             \\
\ce{HSO2}  & Sulfinate             \\
\ce{HSO3}  & Bisulfite             \\
\ce{HOSO}  & Hydroperoxysulfanyl   \\
\ce{HSOO}  & HSOO                  \\
\hline
\end{tabular}
\end{table}

We applied a sensible selection criterion when choosing which sources of the rate constants to use for each reaction. Where possible, we selected rate constants that had been measured experimentally, rather than theoretically. This network was intended to be able to apply to hot Jupiters, whose temperatures can reach several thousand Kelvin. As such, we aimed to pick rate constants that had been measured over a range of temperatures, ideally to temperatures that are realistic for a hot Jupiter, i.e., $2000\,\mathrm{K}$. Unfortunately, many reactions have only been measured at room temperature, resulting in currently unavoidable limitations to the accuracy of these rate constants. Finally, we in general picked the more recently measured rate constants, although many have not had new measurements in many decades. In some cases, the rate constants of both the forward and reverse reaction were available. In these cases, we tried to pick the rate constants for the reaction without an energy barrier, and used thermodynamic reversal to obtain the rates for the reverse reaction. We took the estimated rate constants of sulfur allotrope polymerisation from \cite{Moses2002} and several carbon mono-sulphide, \ce{CS}, reactions from the KIDA database (\citealt{Wakelam2012}) \footnote{http://kida.astrophy.u-bordeaux.fr/}. These reactions are thermodynamically reversed using the NASA7 polynomials from Burcat \footnote{http://garfield.chem.elte.hu/Burcat/burcat.html}, with the full method for these reversals detailed in \cite{Hobbs2019}. There are no polynomials for the species \ce{CS2OH}, \ce{H2C3S} and \ce{HSNO}, so we do not reverse the reactions that include these species. We also include the photodissociation cross-sections for 17 reactions involving eleven sulfur species: \ce{S2}, \ce{S3}, \ce{S4}, \ce{H2S}, \ce{SO}, \ce{SO2}, \ce{SO3}, \ce{S2O}, \ce{OCS}, \ce{CS2} and \ce{CH3SH}. The photo-dissociation cross-section, $\Sigma_{i\, \rightarrow \, j}$ ($\mathrm{m^2}$) is equal to $\sigma_{a,i} \times q_{a,i\, \rightarrow \, j}$, where $\sigma_{a,i}$ ($\mathrm{m^2}$) is the absorption cross-section and $q_{a,i\, \rightarrow \, j}$ is the quantum yield. The wavelengths $i = 1\,$ \AA\, and $j = 10000\,$ \AA\, are the range over which we use these cross-sections. The cross-sections for \ce{H2S}, \ce{SO}, \ce{SO2}, \ce{OCS}, \ce{CS2} and \ce{CH3SH} were taken from PhIDrates \footnote{https://phidrates.space.swri.edu/}, the cross-sections for \ce{S3}, \ce{S4} and \ce{S2} from the MPI-Mainz-UV-VIS Spectral Atlas
of Gaseous Molecules\footnote{ www.uv-vis-spectral-atlas-mainz.org} and the cross-section for \ce{S2} from the Leiden database (\citealt{Heays2017}).

Here we define which parameters are consistent throughout all models, and what changes are applied to specific models. In all of our models:
\begin{itemize}
    \item We set the boundary conditions at the top and bottom of the atmosphere to be zero-flux, such that there is no interaction with the atmosphere outside of our model
    \item When we apply a UV flux to our atmospheres, we choose a mean zenith angle of 57.3$^{\circ}$ (\citealt{Zahnle2008,Hu2012})
    \item We model the planet as if its day-side is always facing the star it orbits, i.e. it's tidally locked, such that it experiences an uninterrupted flux.
\end{itemize}

 Most of our models use a solar metallicity, with values from \cite{Asplund2009}. This gives an elemental ratio, as a fraction of the total number of molecules, of: $X_{\ce{H2}}=0.5\times X_{H}=0.8535$, $X_{\ce{He}}=0.145$, $X_{\ce{C}}=4.584\times10^{-4}$, $X_{\ce{O}}=8.359\times10^{-4}$, $X_{\ce{N}}=1.154\times10^{-4}$, $X_{\ce{S}}=2.250\times10^{-5}$. When we alter the metallicity, we proportionally change the amount of \ce{C}, \ce{O}, \ce{N} and \ce{S} in the atmosphere. \ce{He} is kept constant, and \ce{H2} is altered such that the ratios sum to unity.

There are also parameters that vary between each model. This includes: the pressure-temperature (P-T) profile, the gravity of the planet, the eddy-diffusion profile applied to the model and the spectral irradiance applied to the top of the atmosphere. An overview of these differences can be seen in Table \ref{tab:parameters}.

\begin{table*}
\centering
\caption{Model Parameters}
\label{tab:parameters}
\begin{tabular}{llccccc}
\hline
Case & Analogue Planet & $T$-profile & Gravity & $K_{zz}$-profile & Metallicity & TOA Spectral Irradiance\\
 & & K & cm s$^{-2}$ & cm$^2$ s$^{-1}$ & & photons cm$^{-2}$ s$^{-1}$ \AA$^{-1}$\\
\hline
Figure \ref{fig:Equilibrium1400KSulfur} & Hot Jupiter & 1400 & 1000 & None & Solar & None\\
Figure \ref{fig:Wang17Comp} & Hot Jupiter & 2000 & 1000 & $10^9$ & Solar & $800\times$ Solar\\
Figure \ref{fig:ZahnleWaspComp} & WASP-121 b & \citet{Evans2018} & 875 & $10^9$ & [M/H] = 1.3 & $5300\times$ Solar\\
Figure \ref{fig:IsoEQ} & Hot Jupiter & Multiple & 1000 & None & Solar & None\\
Figure \ref{fig:TripleDEQComp} & Hot Jupiter & Multiple & 1000 & $10^9$ & Solar & $10\times$ Solar\\
Figure \ref{fig:KzzSens} & Hot Jupiter & 1200 & 1000 & Multiple & Solar & $10\times$ Solar\\
Figure \ref{fig:UVSens} & Hot Jupiter & 1200 & 1000 & $10^9$ & Solar & Multiple\\
Figure \ref{fig:HD209SulfurComp} & HD 209458 b & \citet{Hobbs2019} & 936 & \citet{Hobbs2019} & Solar & $800\times$ Solar\\
Figure \ref{fig:51EriSulfurComp} & 51 Eri b & \citet{Moses2016} & 3200 & $10^7$ & Solar & \citet{Moses2016}\\
\hline
\end{tabular}
\end{table*} 
\section{Validation of the Network}
\label{Section3}

In this section we test our model by applying it to the atmospheres of several hot Jupiters. We compare the thermochemistry of our model to an equilibrium solver, FastChem. We also compare and contrast our complete model to a previous sulfur models produced by \cite{Wang2017} and \cite{Zahnle2016}. 

\subsection{Equilibrium Comparison}
\label{Sec: Eq Comp}

\begin{figure}
        \centering
        \includegraphics[width=0.475\textwidth]{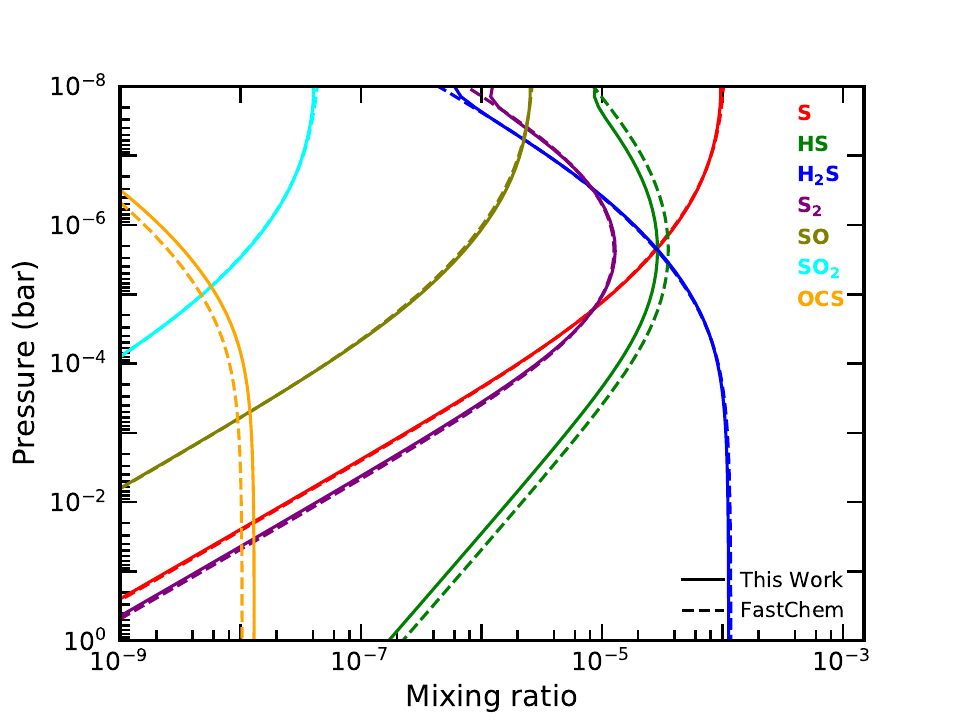}
    \caption[Equilibrium1400KSulfur]{The abundances of significant sulfur bearing species for a purely thermochemical chemistry calculation, taking place in a hot Jupiter atmosphere model that is isothermal at $1400\, \mathrm{K}$, has a gravity of $10\, \mathrm{m\, s^{-2}}$ and a metallicity $5\times$ solar. It has no diffusion or UV spectrum applied to the atmosphere. The solid lines are the output of the model being discussed in this work, while the dashed lines are from the analytical equilibrium solver of FastChem (\citealt{Stock2018}).}
     \label{fig:Equilibrium1400KSulfur}
\end{figure}

As a first step of validating our new sulfur network, we chose to consider our results when only allowing thermochemistry, with no disequilibrium chemistry. We compare to the analytical equilibrium output of FastChem, a chemical equilibrium solver produced by \cite{Stock2018}. In this comparison, we chose to compare a hot Jupiter atmosphere model with an isothermal temperature of $1400\, \mathrm{K}$. The results of this are shown in Figure \ref{fig:Equilibrium1400KSulfur}, where we display only the important sulfur species for comparison. We have a near perfect match compared to the analytic output for every species except for \ce{HS}, which we slightly under-produce, and \ce{OCS}, which we slightly over-produce. 

We expect the causes for these slight dissimilarities to be due to differences in the thermochemical constants we use to reverse our reaction rates. We use the NASA7 polynomials from Burcat \footnote{http://garfield.chem.elte.hu/Burcat/burcat.html} to calculate our Gibbs free energy, while \cite{Stock2018} drew theirs from thermochemical databases, e.g., \cite{Chase1998}. 

Overall, it can be seen that the thermochemistry of our sulfur network is an excellent match to the analytical solution of FastChem. The slight differences seen between the two models are not large enough to significantly impact the chemistry occurring in the atmosphere. Additionally, at a practical level, the difference seen between the models would not be detectable with present observations.

\subsection{Comparison with previous sulfur networks}

The model of \cite{Wang2017} was used to create synthetic spectra to determine whether JWST could detect sulfur and phosphorous species in the atmosphere of giant planets. We make comparisons between our new network and the output of their model for sulfur species. 

\begin{figure}
        \centering
        \includegraphics[width=0.475\textwidth]{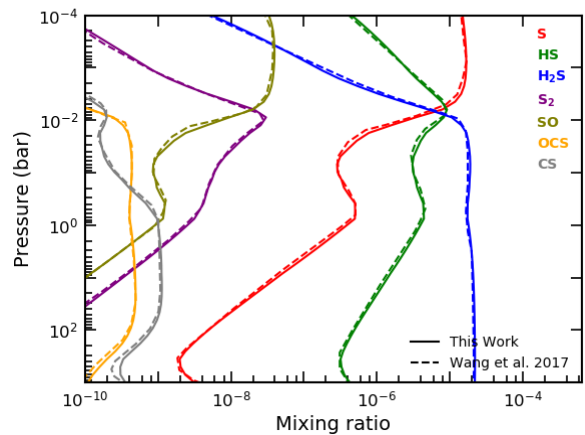}
    \caption[Wang17Comp]{A comparison between our network (solid lines) and one presented in \cite{Wang2017} (dashed lines) for a hot Jupiter model. This atmosphere has an equilibrium temperature of $2000\, \mathrm{K}$, a gravity of $10\, \mathrm{m\, s^{-2}}$, $K_{zz} = \mathrm{10^9\, cm^2\, s^{-1}}$ and with a solar metalicity.}
     \label{fig:Wang17Comp}
\end{figure}

In Figure \ref{fig:Wang17Comp}, the comparison is made for a hot Jupiter model with an atmosphere at an equilibrium temperature of $1400\, \mathrm{K}$, a $K_{zz} = \mathrm{10^9\, cm^2\, s^{-1}}$. In this figure, most of the atmosphere shown is dominated by the effects of thermochemistry, with only the very top possibly being effected by the disequilibrium effects of diffusion. We compare the abundances of the seven most abundant sulfur species, \ce{S}, \ce{S2}, \ce{HS}, \ce{H2S}, \ce{SO}, \ce{OCS} and \ce{CS}. 

Throughout the entire atmosphere we see an excellent match between our own model and that of \cite{Wang2017}, with the abundance of every species matching very closely between the models. With this, we have firmly validated the thermochemistry of our new sulfur network. \cite{Wang2017} did not include considerations of photochemistry in their model, and so further tests are required to validate the sulfur photochemistry in our model. 

\begin{figure}
        \centering
        \includegraphics[width=0.475\textwidth]{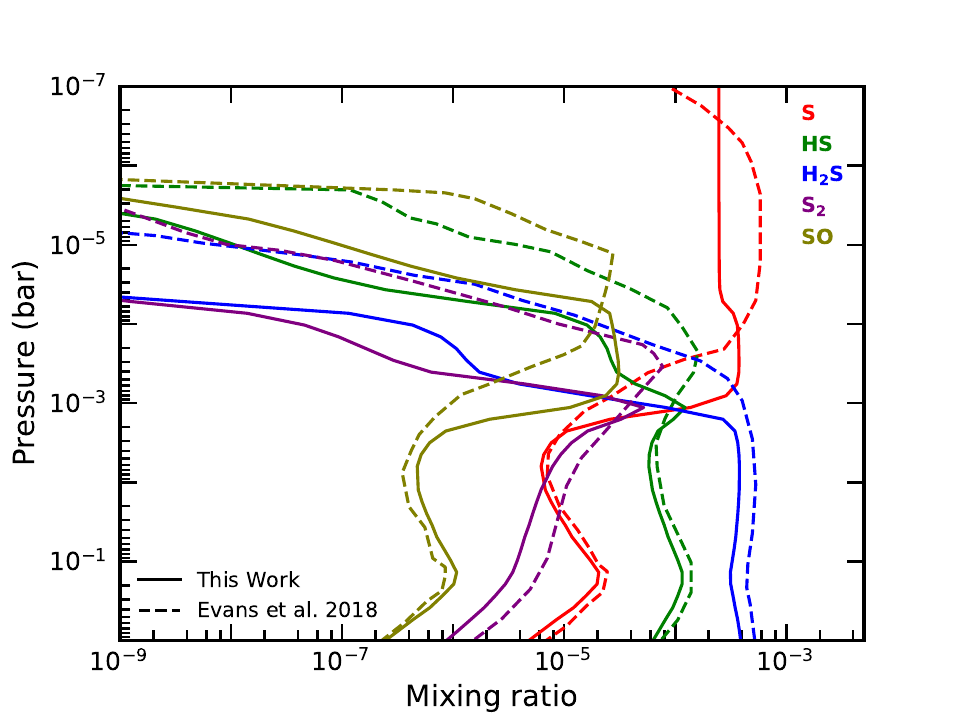}
    \caption[ZahnleWaspComp]{A comparison between our network (solid lines) and the one presented in \cite{Evans2018} (but created by the network of \cite{Zahnle2016}) (dashed lines) for a model of WASP-121 b. The P-T profile for this model is taken from \cite{Evans2018}, with constant $K_{zz} = \mathrm{10^9\, cm^2\, s^{-1}}$, a gravity of $8.75\, \mathrm{m\, s^{-2}}$ and a metalicity $20\times$ solar. The UV spectrum applied to the top of the atmosphere is $5300\times$ Earh's insolation.}
     \label{fig:ZahnleWaspComp}
\end{figure}

In Figure \ref{fig:ZahnleWaspComp} we show our model for the hot Jupiter WASP-121 b, and compare to the abundance profile presented in \cite{Evans2018}, which was produced by the model of \cite{Zahnle2016}. This hot Jupiter is modelled with a metalicity of $20\times$ solar and constant $K_{zz} = \mathrm{10^9\, cm^2\, s^{-1}}$. The model in \cite{Evans2018} used an F6V host star, whereas here we use a solar host star, with its spectrum scaled as if it were the size and temperature of WASP-121. 

We can broadly split the atmosphere in this figure into two regimes. The first is deep in the atmosphere where the pressure is greater than $\mathrm{10^{-3}\,bar}$, and where thermochemical effects are expected to be more important. The second is at pressures less than $\mathrm{10^{-3}\,bar}$ where we expect disequilibrium effects such diffusion and photochemistry to dominate.

In the deep atmosphere, Figure \ref{fig:ZahnleWaspComp} shows that our model has close agreement with the model of \cite{Evans2018}, with a difference in predicted abundances of less than 50\% between the models. We observe a systematic under-abundance of \ce{S}-bearing species in our model compared with that of \cite{Evans2018}. As the non-\ce{S} species match closely, this likely reflects a slightly different atmospheric composition being used. More significant differences can be seen in the upper atmosphere, with most of the displayed species varying by several orders of magnitude between the models. We expect this to be primarily due to the difference in the stellar spectrum being applied to this atmosphere. It can be seen in Figure \ref{fig:ZahnleWaspComp} that the general shape of the abundance profiles are very similar, except that our model predicts photochemistry to begin deeper into the atmosphere, a result consistent with the stronger UV spectrum of a G2V star penetrating further. The only other major difference seen in this figure is the profile of \ce{S} in the upper atmosphere. We believe this to be as a result of differences in how the models treat molecular diffusion.

Our new network has also been used in \cite{Rimmer2021} to model sulfur chemistry on Venus. The model in that work produces results very similar to the observations of Venus's atmosphere, suggesting that the network is valid down to temperatures in the middle atmosphere of Venus; as low as $200\, \mathrm{K}$.

In conclusion, after comparing our new sulfur model against the sulfur models from \cite{Wang2017} and \cite{Evans2018}, we see both similarities and differences in the way we treated thermochemistry and photochemistry. We are confident in the accuracy of our thermochemistry after comparisons with the chemical equilibrium solver in Section \ref{Sec: Eq Comp}. Also, our thermochemical results are very similar to those seen in \cite{Evans2018}. Comparing photochemistry is more difficult, but overall the structure of the abundance profiles suggests the photochemistry between this work and \cite{Evans2018} is similar. The discrepancies that can be seen are likely due to differences in UV cross-sections and stellar spectrum.

\section{Sulfur in Hot Jupiters} 
\label{Section4}

In this section, we investigate sulfur chemistry in the atmosphere of hot Jupiters. We identify the major sulfur species throughout the atmosphere, and note how and why the abundance of these species change. We consider this change in relation to both the planet's temperature and other properties such as diffusion.

\subsection{Local Thermochemical Equilibrium}

To begin with, we investigate the simplest scenario for sulfur chemistry in hot Jupiters: An isothermal atmosphere model that is purely in local thermochemical equilibrium, without any form of diffusion or photochemistry. 

\begin{figure}
        \centering
        \includegraphics[width=0.475\textwidth]{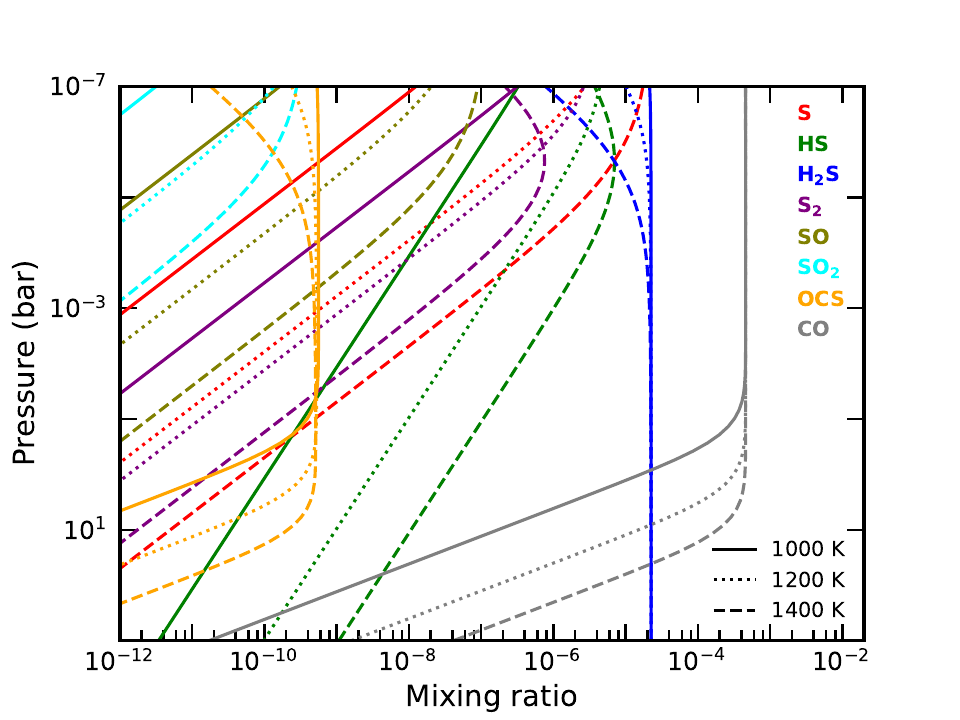}
    \caption[Isothermal Equilibrium]{Sulfur chemistry in three isothermal ($1000 \mathrm{K}$, $1200 \mathrm{K}$ and $1400 \mathrm{K}$) hot Jupiter atmospheres in local thermodynamic equilibrium. This atmosphere has a solar metalicity and a gravity of $10\, \mathrm{m\, s^{-2}}$. It has no diffusion or photochemistry applied to it.}
     \label{fig:IsoEQ}
\end{figure}

In Figure \ref{fig:IsoEQ} we show comparisons of three isothermal atmospheres in chemical equilibrium. The isothermal temperatures of these atmospheres are $1000\,\mathrm{K}$ (solid), $1200\,\mathrm{K}$ (dotted) and $1400\,\mathrm{K}$ (dashed). All three have a solar metalicity and a gravity of $10\, \mathrm{m\, s^{-2}}$. We plot carbon monoxide, \ce{CO}, and sulfur bearing species that have a mixing ratio greater than $10^{-10}$ at any point in the atmosphere. In this manner we can show the full scope of sulfur chemistry throughout the entire atmosphere. 

The first point of note is the preference of sulfur to form \ce{H2S}. At all temperatures shown, \ce{H2S} is the primary carrier of sulfur throughout the majority of the atmosphere, with an abundance of ~$2\times10^{-5}$ for most of the deep atmosphere. It is only at low pressures along the hottest isotherms where this changes, with the primary sulfur carrier becoming atomic \ce{S}. The reactions that set these abundances are

\begin{equation}
\begin{aligned}
\ce{S + H2 &$\, \rightarrow \, $ H + HS}  \\
\ce{HS + H2 &$\, \rightarrow \, $ H + H2S}  \\ \label{reac:Smain}
\ce{2H + M &$\, \rightarrow \, $ H2 +M}  \\
\mathbf{Net:}\ \ce{S + H2 &$\, \rightarrow \, $ H2S}.
\end{aligned}
\end{equation}

High temperatures favour the left hand side of this net equation, preferring to dissociate \ce{H2S} to its constituent \ce{H2} and \ce{S}. In comparison, high pressures favour the formation of \ce{H2S} from \ce{H2} and \ce{S}. This is why \ce{S} becomes the most abundant sulfur molecule only at very low pressures along the high temperature isotherms. 

The mercapto radical involved in this reaction, \ce{HS}, is also of interest, as it acts as an intermediary between \ce{S} and \ce{H2S}. \ce{HS} forms a steady pool that quickly reacts to either \ce{S} or \ce{H2S}, depending on which is more thermodynamically favoured. Generally, the higher temperature atmospheres favour \ce{HS}. In all three models, lower pressure increases the abundance of \ce{HS}. This is unsurprising since, as previously mentioned, the higher temperature and lower pressure favour the dissociation of the large pool of \ce{H2S}, creating more \ce{HS} as a result. Although, at the very top of the atmosphere, above $\mathrm{10^{-7}\,bar}$, this changes. Here, the abundance of \ce{HS} in the $1400\, \mathrm{K}$ atmosphere decreases with pressure and drops below the abundance of \ce{HS} in the $1200\, \mathrm{K}$ atmosphere. This result arises from atomic \ce{S} now being the primary sulfur carrier, since its production is favoured by both temperature and pressure. With the majority of sulfur now being in \ce{S}, the equilibrium abundance of \ce{HS} decreases.

Both sulfur oxides, \ce{SO} and \ce{SO2}, follow similar patterns throughout the atmosphere. They increase in abundance with decreasing pressure, and are more abundant in higher temperature atmospheres. There are two dominant production pathway for \ce{SO}. At high pressures in the lower temperature atmospheres, the pathway is

\begin{equation}
\begin{aligned}
\ce{H2O + M &$\, \rightarrow \, $ OH + H + M}\\
\ce{H2S + H &$\, \rightarrow \, $ H2 + HS}  \\
\ce{HS + H &$\, \rightarrow \, $ S + H2}  \\
\ce{S + OH &$\, \rightarrow \, $ H + SO} \\
\mathbf{Net:}\ \ce{H2S + H2O &$\, \rightarrow \, $ 2H2 + SO},
\end{aligned}
\end{equation}
while in the higher temperature atmosphere, or in any of the modelled atmospheres at low pressures, the pathway is

\begin{equation}
\begin{aligned}
\ce{H2O + M &$\, \rightarrow \, $ OH + H + M}\\
\ce{OH + H &$\, \rightarrow \, $ O + H2}\\
\ce{H2S + O &$\, \rightarrow \, $ H + HSO}  \\
\ce{HSO  &$\, \rightarrow \, $ H + SO} \\
\ce{H + H + M &$\, \rightarrow \, $ H2 + M}  \\
\mathbf{Net:}\ \ce{H2S + H2O &$\, \rightarrow \, $ 2H2 + SO}.
\end{aligned}
\end{equation}

In both pathways, the net reaction is unchanged, with the only differences being in the route taken in the pathway.

For \ce{SO2} the pathways are
\begin{equation}
\begin{aligned}
\ce{H2O + H &$\, \rightarrow \, $ OH + H}\\
\ce{SO + OH &$\, \rightarrow \, $ HOSO }  \\
\ce{HOSO &$\, \rightarrow \, $ SO2 + H } \\
\ce{H + H + M &$\, \rightarrow \, $ H2 + M}  \\
\mathbf{Net:}\ \ce{SO + H2O &$\, \rightarrow \, $ SO2 + H2},
\end{aligned}
\end{equation}
for the deep atmosphere, or more directly
\begin{equation}
\begin{aligned}
\ce{H + H2O &$\, \rightarrow \, $ OH + H2} \\
\ce{SO + OH &$\, \rightarrow \, $ SO2 + H } \\
\mathbf{Net:}\ \ce{SO + H2O &$\, \rightarrow \, $ SO2 + H2},
\end{aligned}
\end{equation}
for the upper atmosphere.

The sulfur molecules \ce{HSO} and \ce{HOSO} referenced in these pathways are not shown in Figure \ref{fig:IsoEQ} due to their very low abundance. The production of both \ce{SO} and \ce{SO2} starts with the destruction of \ce{H2S} by an oxygen radical, \ce{O}. This produces the short lived species \ce{HSO} that quickly dissociates into \ce{SO}. \ce{SO} subsequently reacts with the hydroxyl radical, \ce{OH}. At pressures greater than $\mathrm{10^{-1}\,bar}$, this reaction forms the very short lived species \ce{HOSO}, that then dissociates into \ce{SO2}. At pressures less than $\mathrm{10^{-1}\,bar}$, \ce{SO} and \ce{OH} react directly to form \ce{SO2}. The production of \ce{SO} is significantly faster than the reaction rate of \ce{SO} with \ce{OH}. As a result, the equilibrium abundance of \ce{SO} is much larger than that of \ce{SO2}. Near the top of the atmosphere, above $\mathrm{10^{-6}\,bar}$ in the $1400\, \mathrm{K}$ isothermal atmosphere, the abundances of \ce{SO} and \ce{SO2} stop increasing with decreasing pressure. This happens due to the \ce{H2S} abundance dropping. It is now no longer as available to start the production pathway of \ce{SO} and \ce{SO2}. 

The pathway producing the sulfur molecule \ce{OCS} is

\begin{equation}
\begin{aligned}
\ce{H2S + H  &$\, \rightarrow \, $ HS + H2} \\
\ce{HS + CO &$\, \rightarrow \, $ OCS + H} \\
\mathbf{Net:}\ \ce{H2S + CO &$\, \rightarrow \, $ OCS + H2}.
\end{aligned}
\label{reac:OCS}
\end{equation}

In the deep atmosphere, below $\mathrm{10^{-1}\,bar}$, the abundance of \ce{OCS} is controlled by the abundance of \ce{CO}. Above this pressure, there is a zone for which both \ce{H2S} and \ce{CO} have a constant abundance. This leads \ce{OCS} to stay at a constant abundance at these pressures. Above $\mathrm{10^{-4}\,bar}$ for the $1400\, \mathrm{K}$ atmosphere, and $\mathrm{10^{-6}\,bar}$ for the $1200\, \mathrm{K}$ atmosphere, the drop in the abundance of \ce{H2S} results in the mixing ratio of \ce{OCS} decreasing at a similar rate.

The allotrope of sulfur, \ce{S2}, is produced by the pathway

\begin{equation}
\begin{aligned}
\ce{H2S + H  &$\, \rightarrow \, $ HS + H2} \\
\ce{HS + S &$\, \rightarrow \, $ S2 + H} \\
\mathbf{Net:}\ \ce{H2S + S &$\, \rightarrow \, $ S2 + H2}.
\end{aligned}
\end{equation}

Thus, the abundance of \ce{S2} is controlled by the abundance of \ce{S} and \ce{HS} (and therefore \ce{H2S}). This results in a general increase of abundance as pressure decreases. The exception is for the case of the $1400\, \mathrm{K}$ isotherm, where the drop in abundance of \ce{H2S} and \ce{HS} above $\mathrm{10^{-6}\,bar}$ leads to a decrease in the mixing ratio of \ce{S2}.

\subsection{Disequilibrium}

The atmospheres of most planets are not in chemical equilibrium at low pressure (< $\mathrm{10^{-3}\,bar}$). Therefore it is important to consider how the disequilibrium effects of diffusion and photochemistry can impact the expected sulfur chemistry in a planet's atmosphere. The lowest pressure shown for our disequilibrium figures is $\mathrm{10^{-7}\,bar}$. This is theoretically high enough in the atmosphere that sulfur-ionizing photons could form an ionosphere. The consideration of ion-chemistry is outside the scope of this work, but it is important to note that the top of our modelled atmospheres do not include this effect.

\begin{figure}
        \centering
        \includegraphics[width=0.475\textwidth]{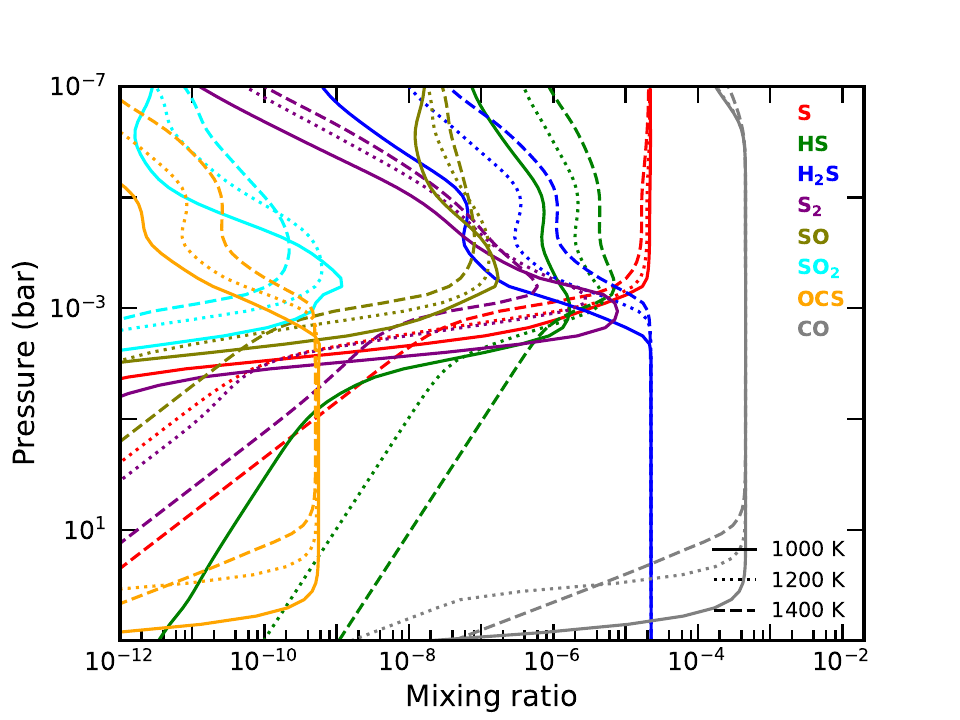}
    \caption[Isothermal Disequilibrium]{Sulfur chemistry in three isothermal ($1000 \mathrm{K}$, $1200 \mathrm{K}$ and $1400 \mathrm{K}$) hot Jupiter atmospheres in steady-state. This atmosphere has a solar metalicity, a gravity of $10\, \mathrm{m\, s^{-2}}$, constant $K_{zz} = \mathrm{10^9\, cm^2\, s^{-1}}$ and a UV flux $10\times$ Earth's.}
     \label{fig:TripleDEQComp}
\end{figure}

In Figure \ref{fig:TripleDEQComp} we show the same three isothermal atmospheres as in the previous section: Isotherms of $1000\,\mathrm{K}$ (Solid), $1200\,\mathrm{K}$ (Dotted) and $1400\,\mathrm{K}$ (Dashed). Once again, all three have a solar metalicity and a gravity of $10\, \mathrm{m\, s^{-2}}$. All of these atmospheres now have a constant eddy diffusion $K_{zz} = \mathrm{10^9\, cm^2\, s^{-1}}$. They also have a UV spectrum applied to the top of the atmosphere of $10\times$ Earth's received solar flux.

In the deep atmosphere, below about $\mathrm{10^{-1}\,bar}$, the abundances are near identical to the equilibrium results presented in the previous section. At this depth, the high pressure ensures that the reaction rates are fast enough that the disequilibrium effects of diffusion and photo-dissociation do not have a significant impact on the chemistry previously presented. 

Higher in the atmosphere, above $\mathrm{10^{-3}\,bar}$, we start to see the significance of disequilibrium chemistry. Many of the sulfur molecules are susceptible to photo-dissociation. 

Once again \ce{H2S} is of particular note. At $\mathrm{10^{-3}\,bar}$ it is rapidly dissociated by free \ce{H} radicals that are produced by photolysis, and then diffuse down from higher in the atmosphere

\begin{equation}
\begin{aligned}
\ce{H2O +} h\nu \ce{ &$\, \rightarrow \, $ OH + H} \\
\ce{H2 + OH &$\, \rightarrow \, $ H2O + H}\\
\ce{H2S + H &$\, \rightarrow \, $ HS + H2} \\
\ce{HS + H &$\, \rightarrow \, $ S + H2} \\
\mathbf{Net:}\ \ce{H2S &$\, \rightarrow \, $ S + H2}.
\end{aligned}
\end{equation}

This effect occurs at all temperatures. Although, in the colder models, we see the abundance of \ce{H2S} driven much further away from its equilibrium value, by more than two orders of magnitude, compared to the hotter models. This is because of the longer chemical timescales at lower temperatures, which slows the recombination reactions. Overall, the rapid destruction of \ce{H2S} allows for full dissociation to atomic \ce{S}. This results in \ce{S} being the primary sulfur molecule at all temperatures and pressures above $\mathrm{10^{-3}\,bar}$.

The production of a large available pool of \ce{S} has knock-on effects for the abundance of \ce{SO} and \ce{SO2}. Above $\mathrm{10^{-3}\,bar}$ the abundance of \ce{SO} is controlled by the reaction chain of

\begin{equation}
\begin{aligned}
\ce{SO +} h\nu \ce{ &$\, \rightarrow \, $ S + O} \\
\ce{H2O +} h\nu \ce{ &$\, \rightarrow \, $ OH + H} \\
\ce{S + OH  &$\, \rightarrow \, $ SO + H} \\
\ce{H + H + M &$\, \rightarrow \, $ H2 + M} \\
\mathbf{Net:}\ \ce{H2O &$\, \rightarrow \, $ H2 + O},
\end{aligned}
\end{equation}
and \ce{SO2} by a very similar chain
\begin{equation}
\begin{aligned}
\ce{SO2 +} h\nu \ce{ &$\, \rightarrow \, $ SO + O} \\
\ce{H2O +} h\nu \ce{ &$\, \rightarrow \, $ OH + H} \\
\ce{SO + OH  &$\, \rightarrow \, $ SO2 + H} \\
\ce{H + H + M &$\, \rightarrow \, $ H2 + M} \\
\mathbf{Net:}\ \ce{H2O &$\, \rightarrow \, $ H2 + O}.
\end{aligned}
\end{equation}

Overall, both these reaction pathways act as a dissipative cycle that increases the abundance of atomic \ce{O} in the atmosphere, while preventing \ce{SO} and \ce{SO2} from being permanently dissociated. The abundances of both \ce{SO} and \ce{SO2} are primarily determined by the amount of atomic \ce{S} in the atmosphere. As a result the abundances of \ce{SO} and \ce{SO2} are nearly temperature independent throughout the upper atmosphere: \ce{SO} reaches abundances around $\mathrm{10^{-7}}$, while \ce{SO2} is never greater than $\mathrm{10^{-9}}$.

\ce{S2} is significantly affected by the introduction of disequilibrium chemistry. The large increase in both \ce{HS} and \ce{S} around $\mathrm{10^{-3}\,bar}$ greatly boosts its production through the reaction

\begin{equation}
\begin{aligned}
\ce{S + HS  &$\, \rightarrow \, $ S2 + H}.
\end{aligned}
\end{equation}

This is particularly prominent in the lower temperature isotherms. In the $1000\,\mathrm{K}$ atmosphere, the abundance of \ce{S2} reaches nearly $\mathrm{10^{-5}}$ (compared to only $\mathrm{10^{-10}}$ at the same pressure and temperature in the local thermochemical equilibrium case). However, above $\mathrm{10^{-3}\,bar}$, it is rapidly photo-dissociated into 2\ce{S}. The outcome is a quick drop-off in its abundance at all temperatures. 

\ce{OCS} is not greatly affected by the introduction of disequilibrium chemistry. Its abundance is still determined by the pathway shown in equation \ref{reac:OCS}.

\subsection{Diffusion Strength}
\begin{figure}
        \centering
        \includegraphics[width=0.475\textwidth]{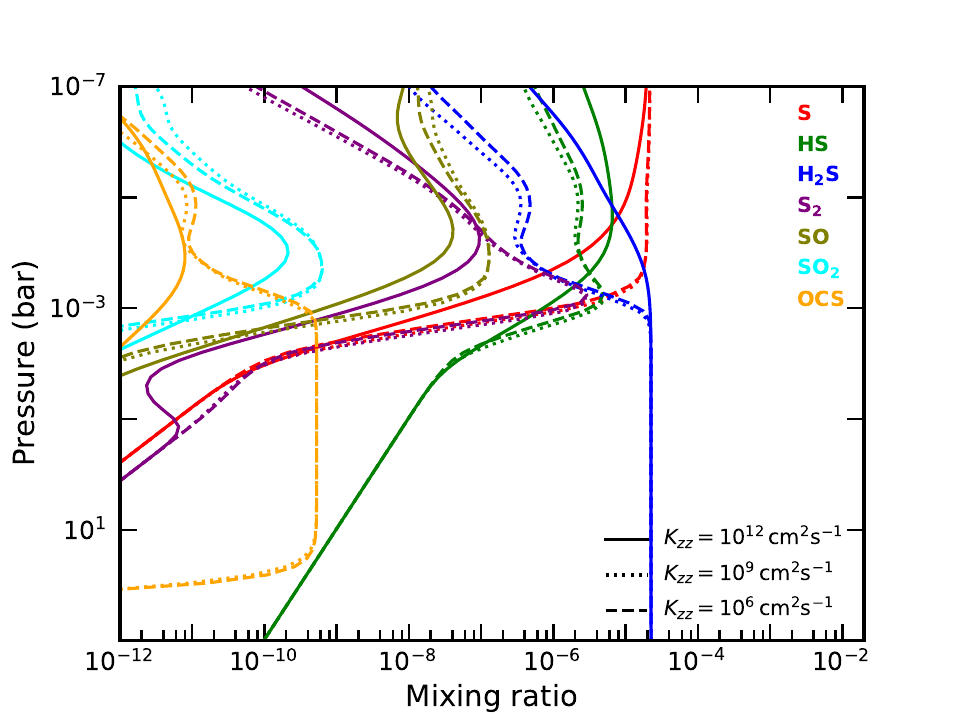}
    \caption[Isothermal Sensitivity]{Sulfur chemistry in a $1200\,\mathrm{K}$ isothermal hot Jupiter atmosphere. This model explores the sensitivity of sulfur chemistry to the strength of the vertical mixing, by comparing models with three different $K_{zz}$ values. This atmosphere has a solar metalicity, a gravity of $10\, \mathrm{m\, s^{-2}}$ and has a UV spectrum $10\times$ Earth's applied to it.}
     \label{fig:KzzSens}
\end{figure}

It is important to understand how our choices for the parameters of the disequilibrium chemistry can affect the overall distribution of sulfur chemistry in planetary atmospheres. To that end, in Figure \ref{fig:KzzSens}, we compare three atmospheric models of a $1200\, \mathrm{K}$ isothermal hot Jupiter with different strengths of eddy-diffusion, $K_{zz}$. We do so to test the sensitivity of sulfur chemistry to the strength of the vertical mixing in the atmosphere. We chose to use two extremes of diffusion, $K_{zz} = \mathrm{10^6\, cm^2\, s^{-1}}$ (Dashed) and $K_{zz} = \mathrm{10^{12}\, cm^2\, s^{-1}}$ (Solid), to understand the limiting cases for diffusion. We also include the previously used average case of $K_{zz} = \mathrm{10^9\, cm^2\, s^{-1}}$ (Dotted) as a comparison.

Immediately seen is that the weakest mixing, $K_{zz} = \mathrm{10^6\, cm^2\, s^{-1}}$, does not differ greatly from the intermediate case of $K_{zz} = \mathrm{10^9\, cm^2\, s^{-1}}$. This suggests that $K_{zz} = \mathrm{10^9\, cm^2\, s^{-1}}$ is already sufficiently low that eddy-diffusion is having little effect upon sulfur chemistry.

For the strongest diffusion, $K_{zz} = \mathrm{10^{12}\, cm^2\, s^{-1}}$, there are several significant differences to the abundance of sulfur molecules. Most of these effects stem from the diffusion quenching the \ce{H2S}, such that it stays as the most abundant sulfur molecule up until $\mathrm{10^{-5}\,bar}$. This results in a lower atomic \ce{S} abundance, which has a knock-on effect of also lowering the abundance of both \ce{SO} and \ce{SO2}. 

The sulfur allotrope \ce{S2} is also significantly affected by the stronger diffusion. Its abundance between $\mathrm{10^{-1}\,bar}$ and $\mathrm{10^{-4}\,bar}$ drops by up to two orders of magnitude compared to the weaker diffusion cases. This is largely caused by the \ce{S2} diffusing higher into the atmosphere where it can be photo-dissociated, which causes the drop-off in abundance from $\mathrm{10^{-1}\,bar}$. Above $\mathrm{10^{-4}\,bar}$, the stronger diffusion lifts the abundance of \ce{S2} faster than it can be dissociated, to mixing ratios slightly above that of the weaker diffusion models.

Overall, the strength of the vertical mixing does not have much impact on most sulfur molecules. This conclusion is in line with previous studies by \cite{Zahnle2009} and \cite{Zahnle2016}, in which they found that vertical mixing was only significant in lower temperatures due to how fast sulfur reactions are. The atmospheres being studied in this work are sufficiently hot that the atmosphere reaches a steady-state balance between the thermo-chemical and photo-chemical elements. The effects of diffusion are thus negated unless the value of $K_{zz}$ is increased to extreme values. At these extremes, the strong mixing may be able to keep \ce{H2S} detectable higher in the atmosphere, but strong mixing also significantly suppresses the abundance of \ce{S2}, preventing its abundance from reaching detectable levels. 

\subsection{UV Strength}

\begin{figure}
        \centering
        \includegraphics[width=0.475\textwidth]{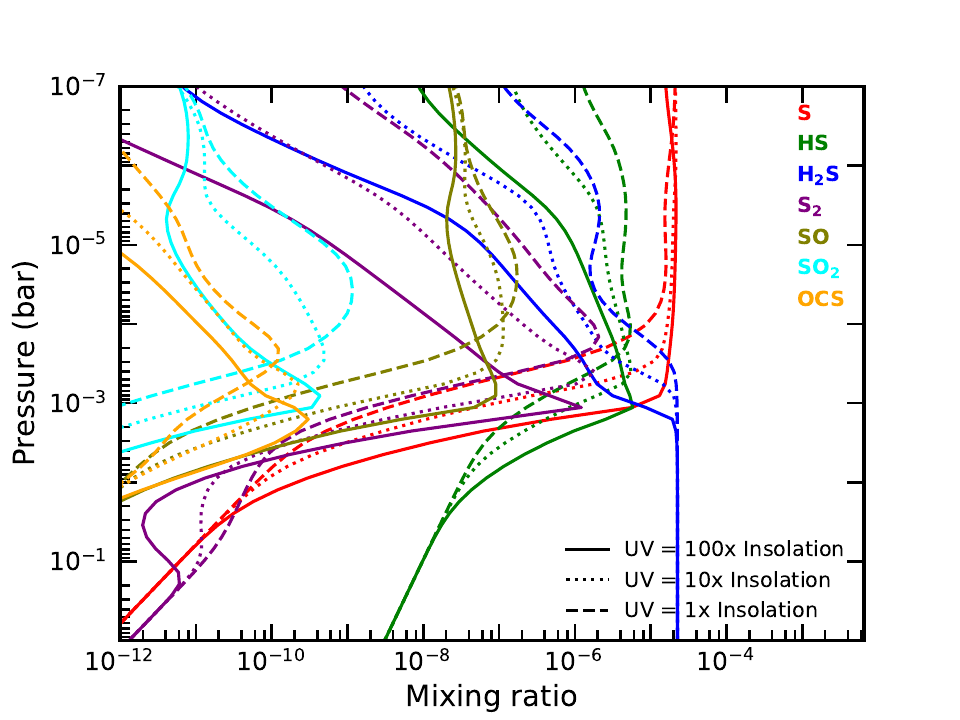}
    \caption[Isothermal Sensitivity]{Sulfur chemistry in a $1200\, \mathrm{K}$ isothermal hot Jupiter atmosphere. This model explores the sensitivity of sulfur chemistry to the strength of the UV flux, by comparing models with three different strength of UV. This atmosphere has a solar metalicity, $K_{zz} = \mathrm{10^{9}\, cm^2\, s^{-1}}$ and a gravity of $10\, \mathrm{m\, s^{-2}}$.}
     \label{fig:UVSens}
\end{figure}

We next examine how altering the strength of the UV flux being applied to the top of the model can alter the sulfur chemistry. While this is not a very realistic change to be made independent of all other parameters, (since the flux strength is most commonly set by the distance of the planet from the star, which would greatly impact the atmosphere's temperature), it is important to understand our network's sensitivity to changes in the rate of photochemistry. As such, we present three different models of a hot Jupiter with a $1200\, \mathrm{K}$ isothermal atmosphere in Figure \ref{fig:UVSens}. Two have extremes of UV irradiation at $100\times$ Earth's (Solid) and $1\times$ Earth's (Dashed). The other has $10\times$ Earth's (Dotted) irradiation, the same as previously used in this section. 

Most species show no change deep in the atmosphere, below the $\mathrm{10^{-2}\,bar}$ level. No species show any difference below $\mathrm{10^{-1}\,bar}$. This shows that even for extreme irradiation, the deepest parts of the atmosphere are difficult to move away from thermochemical equilibrium.

The species primarily affected by the change in UV flux is \ce{H2S}. Its abundance drops by an order of magnitude at $\mathrm{10^{-3}\,bar}$ for the highest flux compared to the lowest flux. This also greatly impacts the \ce{S} at the same pressure. Its abundance is up to five orders of magnitude greater for the greatest flux compared to the smallest. This difference is only seen for a small pressure window however, and the abundance of \ce{S} is very similar across all three fluxes above $\mathrm{10^{-4}\,bar}$. 

In this same pressure window, the greater availability of \ce{S} caused by the stronger flux allows the abundance of both \ce{SO} and \ce{SO2} to grow much larger than in the models with weaker fluxes. However, above $\mathrm{10^{-4}\,bar}$ this is reversed as the available \ce{S} becomes similar in all three models, resulting in the strongest UV model depleting \ce{SO} and \ce{SO2} by about an order of magnitude compared to the weakest UV model. 

The comparative abundance of \ce{S2} goes through several phases. Deep in the atmosphere, around $\mathrm{10^{-1}\,bar}$, the stronger UV flux can penetrate deep enough to begin to dissociate it into 2\ce{S}. However, not far above this, around $\mathrm{10^{-2}\,bar}$, the excess \ce{S} that are produced by \ce{H2S} dissociation in the stronger UV cases are sufficient to boost the production of \ce{S2} above its rate of photo-dissociation. Thus we see more \ce{S2} in the stronger UV model, up until between $\mathrm{10^{-3}\,bar}$ and $\mathrm{10^{-4}\,bar}$ when the availability of atomic \ce{S} equalises across all three models. Above this, the stronger UV flux once again dominates in dissociating \ce{S2}, resulting in its abundance being several orders of magnitude lower than in the weak UV model.

Overall, the strength of the UV flux simply shifts the abundance profiles deeper into the atmosphere. This happens because the stronger flux can penetrate deeper, with an increase of two orders of magnitude in UV flux able to push the profiles of sulfur molecules one order of magnitude deeper in pressure into the atmosphere.  
\section{The impact of Sulfur's inclusion on atmospheric chemistry} 
\label{Section5}

In this section we investigate what effect the sulfur chemistry has upon the modelled composition of both the hot Jupiter HD 209458b and the warm Jupiter 51 Eri b. We compare atmospheric models of these planets with and without sulfur included in our network. In this way we isolate sulfur's impact, and examine the pathways in our network that lead to these chemical differences occurring. 

\subsection{HD 209458 b}

\begin{figure}
        \centering
        \includegraphics[width=0.475\textwidth]{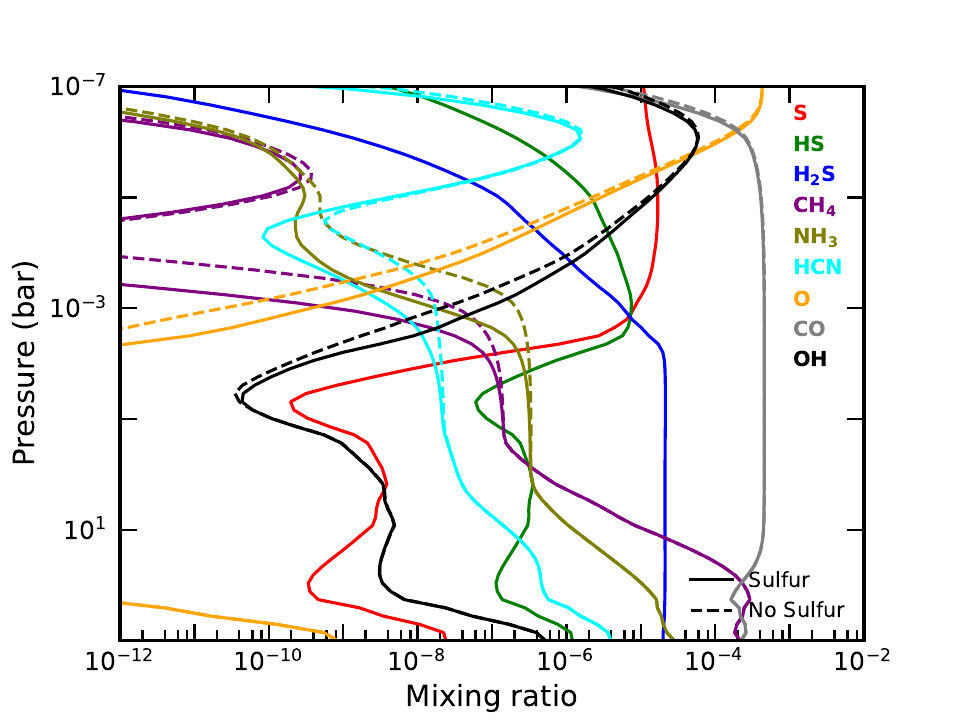}
    \caption[HD209SulfurComp]{A comparison of the abundance profiles for the hot Jupiter HD 209458b both with (solid lines) and without (dashed lines) our new sulfur network being included. This model uses the pressure - temperature profile and $K_{zz}$ profile from \cite{Hobbs2019}. It has solar metalicity, a gravity of $9.36\, \mathrm{m\, s^{-2}}$ and has a UV spectrum $800\times$ Earth's applied to it.}
     \label{fig:HD209SulfurComp}
\end{figure}

We chose to examine the effects of sulfur chemistry on the atmosphere of the well studied hot Jupiter, HD 209458b. We have previously investigated this planet with \textsc{Levi} in \cite{Hobbs2019}, and we return to it again. We do so now with our updated chemical model to discover how the inclusion of sulfur chemistry affects the predicted abundance profiles of this hot Jupiter's atmosphere. 

As can be seen in Figure \ref{fig:HD209SulfurComp}, we see no difference due to the inclusion of sulfur below 1 bar, in the deep atmosphere of the planet. At this depth, the abundance of the main sulfur species is set by the same reaction path as shown in Reaction \ref{reac:Smain}. These reactions are very rapid and interactions with other species are slow by comparison, preventing sulfur from having any effect on CNO chemistry in the deep atmosphere.

Higher in the atmosphere, around $\mathrm{10^{-3}\,bar}$, sulfur has a much more significant effect on the CNO chemistry. In Figure \ref{fig:HD209SulfurComp}, it can be seen that at this pressure the inclusion of sulfur decreases the predicted abundance of \ce{NH3}, \ce{CH4} and \ce{HCN} by up to two orders of magnitude. 

Through a thorough and careful analysis of the fastest reaction pathways that cause the destruction of \ce{NH3}, \ce{CH4} and \ce{HCN}, we can identify how and why sulfur causes these differences. Figure \ref{fig:HD209SulfurComp} shows that sulfur chemistry changes where these three molecules stop being quenched. Molecules stop being quenched when the chemical timescale of the rate limiting step in the fastest destruction pathway becomes faster than the dynamical timescale in that region of the atmosphere. By examining this region for \ce{NH3}, \ce{CH4} and \ce{HCN}, and how it changes location with the inclusion of sulfur, can determine why sulfur chemistry effects the abundance of CNO species within the atmosphere of a hot Jupiter. We use the chemical timescale as
\begin{equation}
\tau_{chem} = \frac{[X]}{d[X]/dt},
\end{equation}
where $[X]$ is the number density of the molecule we are investigating, either \ce{NH3}, \ce{CH4} or \ce{HCN}. $d[X]/dt$ is the change in number density of $X$ with time, which is approximately the reaction rate of the rate limiting step in the fastest destruction pathway of $[X]$. We treat the dynamic timescale as
\begin{equation}
\tau_{dyn} = \frac{H_0 dH}{K_{zz}},
\end{equation}
where $H_0$ is the scale height of the atmosphere, $dH$ is the thickness of one atmospheric layer in our model, and $K_{zz}$ is the eddy-diffusion (\citealt{Visscher2011}).

\begin{figure}
        \centering
        \includegraphics[width=0.475\textwidth]{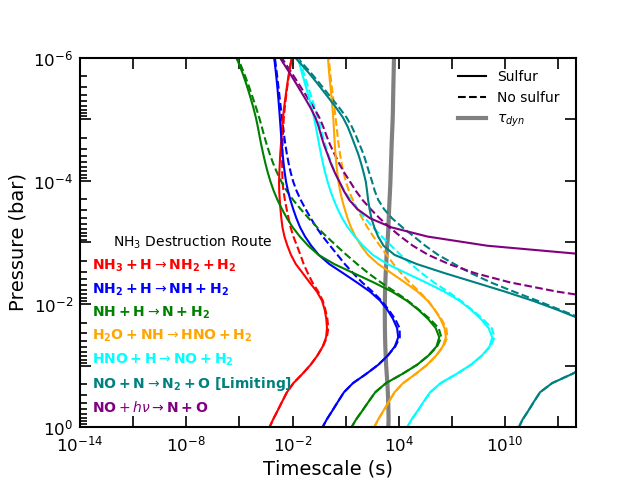}
    \caption[H3NDestruction]{A comparison of the chemical timescales in the most efficient pathway to destroy \ce{NH3} in the atmosphere of HD 209458b, for both an sulfurous and a sulfur-free atmosphere. The dynamic timescale is also shown in this figure. The pressure at which the rate limiting step occurs on a shorter timescale than the dynamic timescale is where we expect to see quenching of \ce{NH3} end in the atmosphere of HD 209458b. The increase in the timescale of the photodissociation of \ce{NO} with the inclusion of sulfur is responsible for the difference in where the rate limiting drops below the dynamic timescale. }
     \label{fig:H3NDestruction}
\end{figure}

The most efficient pathway for the destruction of \ce{NH3} leads the nitrogen to end up in the stable molecule \ce{N2}, via the following scheme:

\begin{equation}
\begin{aligned}
\ce{2 (NH3 + H &$\, \rightarrow \, $ NH2 + H2)} \\
\ce{2 (NH2 + H &$\, \rightarrow \, $ NH + H2)} \\
\ce{NH + H &$\, \rightarrow \, $ N + H2} \\
\ce{NH + H2O &$\, \rightarrow \, $ HNO + H2} \\
\ce{HNO + H &$\, \rightarrow \, $ NO + H2} \\
\boldsymbol{\ce{NO + N }}&\boldsymbol{\ce{$\, \rightarrow \, $ N2 + O}} \\
\ce{O + H2 &$\, \rightarrow \, $ OH + H} \\
\ce{OH + H2 &$\, \rightarrow \, $ H2O + H} \\
\ce{2 (H2 + M &$\, \rightarrow \, $ H + H + M)} \\
\mathbf{Net:}\  \ce{2 NH3 &$\, \rightarrow \, $ N2 + 3H2}, 
\end{aligned}
\end{equation}
with the reaction in bold being the rate limiting step in the pathway. We show the chemical timescales of each of the main steps in this pathway in Figure \ref{fig:H3NDestruction}, as well as the dynamic timescale, all as a function of pressure. We compare these rates for both an atmosphere including sulfur and a sulfur free atmosphere. When we examine where $\tau_{chem}$ of the rate limiting step, \ce{NO + N $\, \rightarrow \, $ N2 + O}, becomes shorter than $\tau_{dyn}$, we see a significant pressure difference between the sulfur and sulfur-free cases being examined. This pressure difference matches up precisely to where we see \ce{NH3} quenching end in Figure \ref{fig:HD209SulfurComp}; around $\mathrm{10^{-3}\,bar}$ with sulfur, and $\mathrm{3x10^{-4}\,bar}$ without sulfur.

The presence of \ce{NO} in the rate limiting step is why the inclusion of sulfur significantly effects its timescale. Figure \ref{fig:H3NDestruction} also includes the timescale of \ce{NO} photodissociation, and it can be seen that the presence of sulfur increases the timescale of this reaction by many orders of magnitude between $\mathrm{10^{-3}}$ and $\mathrm{10^{-2}\,bar}$. This large slowing in the destruction of \ce{NO} results in significantly more \ce{NO} available in the atmosphere, greatly increasing the rate of the rate limiting step, resulting in the difference in the location of the end of quenching between the two models, and the lower abundance of \ce{NH3} seen in the sulfurous model of HD 209458b.

\begin{figure}
        \centering
        \includegraphics[width=0.475\textwidth]{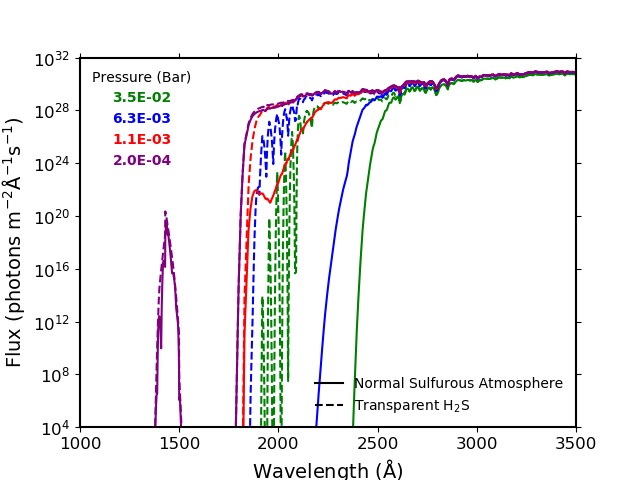}
    \caption[SFluxchnage]{The apparent flux at different pressures within the atmosphere of a sulfurous HD 209458b. To show the effect of sulfur shielding in this atmosphere, we compare the normal flux to the flux if \ce{H2S} was transparent.}
     \label{fig:SFluxchange}
\end{figure}

The difference in the photochemical rate of \ce{NO} arises from the UV shielding by sulfur molecules. In particular, we find that \ce{H2S} is responsible for most of the additional UV shielding when sulfur is introduced to the atmosphere. This can be seen in Figure \ref{fig:SFluxchange} where we show the UV flux at different pressure levels, and compare to the same atmosphere but with transparent \ce{H2S}. We find that \ce{H2S} causes the available flux around $\mathrm{10^{-3}\,bar}$ to drop by tens of orders of magnitude, depending on the wavelengths being examined. Thus \ce{H2S} blocks a large fraction of the incoming UV light, allowing \ce{NO} to survive until higher in the atmosphere compared to a sulfur free atmosphere.

When we examine the dominant destruction pathway for \ce{CH4}, we find that the carbon tends to end up in \ce{CO} via;

\begin{equation}
\begin{aligned}
\ce{CH4 + H &$\, \rightarrow \, $ CH3 + H2} \\
\boldsymbol{\ce{CH3 + OH }}&\boldsymbol{\ce{$\, \rightarrow \, $ CH2O + H2}} \\
\ce{CH2O + H &$\, \rightarrow \, $ CHO + H)} \\
\ce{CHO + H &$\, \rightarrow \, $ CO + H2} \\
\ce{H2O + H &$\, \rightarrow \, $ OH + H2} \\
\ce{2 (H2 + M &$\, \rightarrow \, $ H + H + M)} \\
\mathbf{Net:}\  \ce{CH4 + H2O &$\, \rightarrow \, $ CO + 3H2}, 
\end{aligned}
\end{equation}
with the bold reaction once again being the rate limiting step in the this reaction pathway. We show the main timescales of this reaction pathway in Figure \ref{fig:CH4Destruction}. Once again we see the rate limiting step's timescale become smaller than the dynamic timescale at different pressures in a sulfurous vs. a non-sulfurous atmosphere. The explanation for this lies within a different set of reactions: 

\begin{equation}
\begin{aligned}
\ce{SO +}\ h\nu \ce{&$\, \rightarrow \, $ S + O} \\
\ce{O + H2 &$\, \rightarrow \, $ OH + H} \\
\ce{S + OH &$\, \rightarrow \, $ SO + H}.
\end{aligned}
\label{reac:SO}
\end{equation}

In a sulfurous atmosphere, the photodissociation of \ce{SO} produces \ce{O} significantly faster than any other reaction between $\mathrm{10^{-3}}$ and $\mathrm{10^{-4}\,bar}$. This results in a corresponding rate increase in the reaction \ce{O + H2 \, \rightarrow \, OH + H}. The timescales of both of these reactions are also shown in Figure \ref{fig:CH4Destruction}. The result of these rate increases can be seen in Figure \ref{fig:HD209SulfurComp}, where both \ce{O} and \ce{OH} have their abundances increased by at least an order of magnitude around $\mathrm{10^{-3}\,bar}$. The rate limiting step in the destruction of \ce{CH4} is dependant on the \ce{OH} abundance. Thus, the consequence of an increase in \ce{OH} abundance is that the timescale of the rate limiting step in a sulfurous atmosphere becomes shorter than the dynamic timescale deeper in the atmosphere. This leads to the quenching of \ce{CH4} ending deeper in the atmosphere, and the difference in abundance between sulfurous and non-sulfurous atmosphere, as seen in Figure \ref{fig:HD209SulfurComp}.

\begin{figure}
        \centering
        \includegraphics[width=0.475\textwidth]{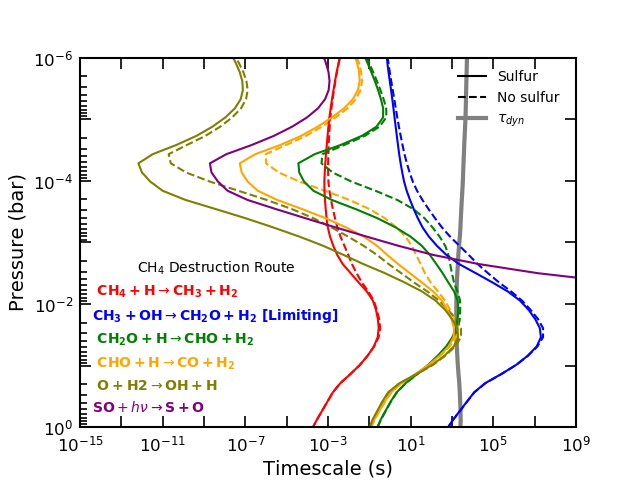}
    \caption[CH4Destruction]{A comparison of the chemical timescales in the most efficient pathway to destroy \ce{CH4} in the atmosphere of HD 209458b, for both an sulfurous and a sulfur-free atmosphere. The dynamic timescale is also shown in this figure. The pressure at which the rate limiting step occurs on a shorter timescale than the dynamic timescale is where we expect to see quenching of \ce{CH4} end in the atmosphere of HD 209458b. The photodissociation of \ce{SO} causes a 30x increase in available \ce{O}, which greatly increases the rate that \ce{O} becomes \ce{OH}, and thus is responsible for the change in the rate limiting step of this reaction pathway.}
     \label{fig:CH4Destruction}
\end{figure}

We have found the most efficient pathway for the destruction of \ce{HCN} to be via the route:

\begin{equation}
\begin{aligned}
\ce{HCN + M &$\, \rightarrow \, $ HNC + M} \\
\boldsymbol{\ce{HNC + OH }}&\boldsymbol{\ce{$\, \rightarrow \, $ HNCO + H}} \\
\ce{HNCO + H &$\, \rightarrow \, $ CO + NH2} \\
\ce{NH2 + H2 &$\, \rightarrow \, $ NH3 + H} \\
\ce{H2O + H &$\, \rightarrow \, $ OH + H2} \\
\mathbf{Net:}\  \ce{HCN + H2O &$\, \rightarrow \, $ CO + NH3}, 
\end{aligned}
\end{equation}
with the rate limiting step in bold. The explanation for the difference in abundance of \ce{HCN} in sulfurous and non-sulfurous atmospheres follows the same reasoning as for \ce{CH4}. Once again, the rate limiting step contains \ce{OH}, and thus Reactions \ref{reac:SO} explain why the inclusion of sulfur increases the abundance of \ce{OH}, and thus the increase the rate at which \ce{HCN} is destroyed.

\begin{figure}
        \centering
        \includegraphics[width=0.475\textwidth]{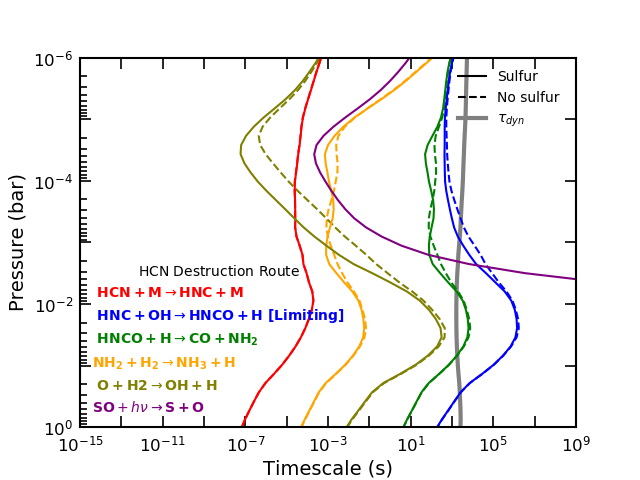}
    \caption[HCNDestruction]{A comparison of the chemical timescales in the most efficient pathway to destroy \ce{HCN} in the atmosphere of HD 209458b, for both an sulfurous and a sulfur-free atmosphere. The dynamic timescale is also shown in this figure. The pressure at which the rate limiting step occurs on a shorter timescale than the dynamic timescale is where we expect to see quenching of \ce{HCN} end in the atmosphere of HD 209458b. The photodissociation of \ce{SO} causes a 30x increase in available \ce{O}, which greatly increases the rate that \ce{O} becomes \ce{OH}, and thus is responsible for the change in the rate limiting step of this reaction pathway.}
     \label{fig:CH4Destruction}
\end{figure}

Overall, the change in the abundance of both \ce{HCN} and \ce{CH4} with the inclusion of sulfur show the catalytic role that sulfur chemistry can have in planetary atmospheres (as also noted by \citealt{Zahnle2016})

Figure \ref{fig:HD209SulfurComp} shows that, at pressures lower than $\mathrm{10^{-5}\,bar}$, the effects of sulfur once again become insignificant.
Above this height, \ce{H2S} begins to drop in abundance, making the shielding of \ce{NO} less effective. This results in the sulfurous and non-sulfurous photodissociation rates of \ce{NO} approach similar rates once more, such that the overall destruction rate of \ce{NH3} is now approximately the same in both models. Additionally, above this height, \ce{SO} is no longer the fastest producer of \ce{O}, with other reactions such as \ce{OH} photodissociation becoming more dominant. As such, both sulfurous and non-sulfurous destruction rates of \ce{CH4} and \ce{HCN} become similar. In consequence, the very upper atmospheres of hot Jupiters that contain sulfur look nearly indistinguishable from those without sulfur.

There is potential for detecting the differences caused by the inclusion of sulfur in hot Jupiters' atmospheres. There has already been evidence for \ce{CH4} (\citealt{Guilluy2019}) and \ce{HCN} (\citealt{Hawker2018,Cabot2019}) detections in the atmosphere's of hot Jupiters. Most retrievals identify species, and their abundances, around the pressure of $\mathrm{10^{-3}\,bar}$. Thus, it should be possible to detect the drop in abundance of these molecules due to sulfur. However, the pressure range over which there is an identifiable difference due to sulfur is rather limited. Any investigated spectra of this planet would need to be very precise, more so than any currently obtainable.

\subsection{Warm Jupiters}

Previously, there has been discussion in the literature about how sulfur on warm Jupiters may affect the abundance of species in the atmosphere of these planets. Of particular note is the warm Jupiter 51 Eri b, which both \cite{Moses2016} and \cite{Zahnle2016} have modelled in the past. \cite{Moses2016} ran models without any sulfur in the atmosphere, while \cite{Zahnle2016} ran models with sulfur. Both works studied the chemistry of 51 Eri b in great detail, both with and without sulfur. What is of particular interest though are the differences found between the two models. As \cite{Moses2016} notes, they find up to two orders of magnitude more \ce{CO2} than \cite{Zahnle2016} for otherwise identical conditions. Thus, our focus in this section will be on examining the warm Jupiter 51 Eri b with both our sulfurous and non-sulfurous models to determine if sulfur is responsible for this difference, and, if it is, why.

\begin{figure}
        \centering
        \includegraphics[width=0.475\textwidth]{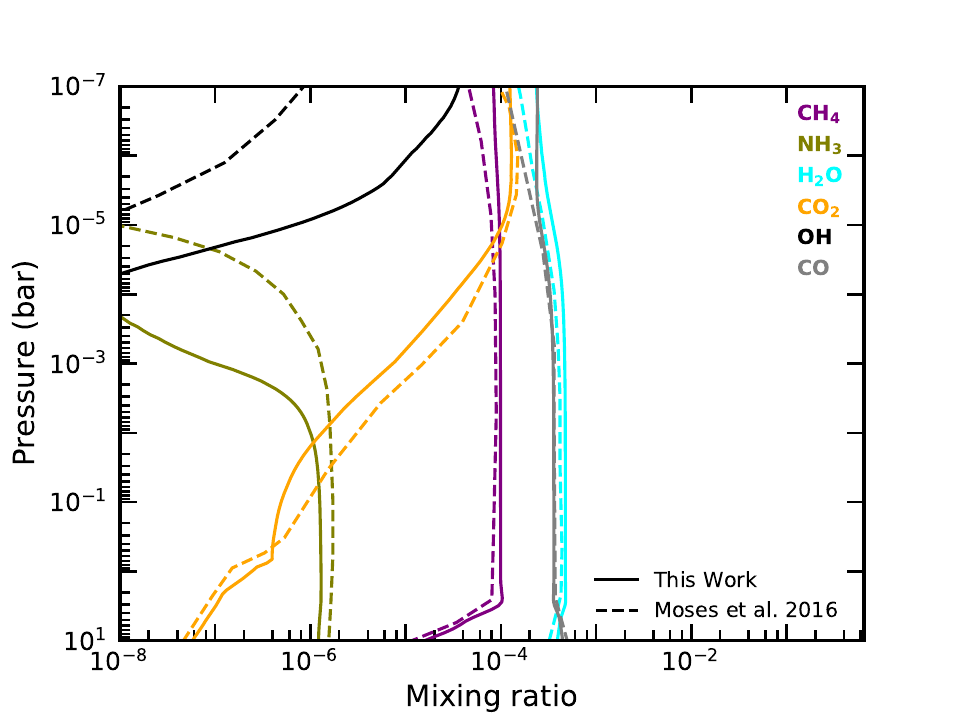}
    \caption[51Moses]{A comparison of the abundance profiles for the warm Jupiter 51 Eri b. The solid lines are the sulfur free model from this work and the dashed lines are the model from \cite{Moses2016}. We used the temperature profile, $K_{ZZ}$ profile and UV flux profile from \cite{Moses2016}. The planet has a gravity of $32\, \mathrm{m s^{-2}}$ and a solar metallicity.}
     \label{fig:51Moses}
\end{figure}

\begin{figure}
        \centering
        \includegraphics[width=0.475\textwidth]{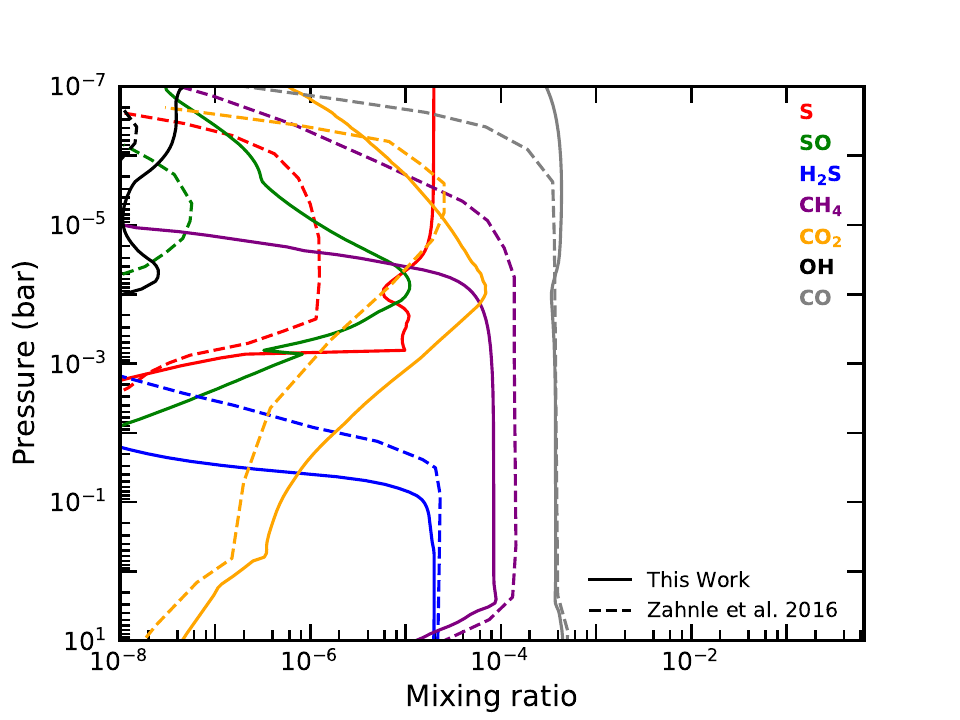}
    \caption[51Moses]{A comparison of the abundance profiles for the warm Jupiter 51 Eri b. The solid lines are the sulfurous model from this work and the dashed lines are the model from \cite{Zahnle2016}. We used the temperature profile and UV flux profile from \cite{Moses2016}, and used a constant $K_{zz} = \mathrm{10^7\, cm^2\, s^{-1}}$. The planet has a gravity of $32\, \mathrm{m\, s^{-2}}$ and a solar metallicity.}
     \label{fig:51Zahnle}
\end{figure}

 To begin this section, we directly compare our model for the atmosphere of 51 Eri b to that of \cite{Moses2016} and \cite{Zahnle2016} as a benchmark for whether differences in the models themselves could be responsible for the difference in \ce{CO2} abundance. In Figure \ref{fig:51Moses} we compare our sulfur free model to the model from \cite{Moses2016}, using the pressure temperature profile, $K_{zz}$ profile and stellar UV flux also from \cite{Moses2016}. In Figure \ref{fig:51Zahnle} we compare our sulfurous model to the model from \cite{Zahnle2016}. We use the same temperature profile and UV flux, but with a constant $K_{zz} = \mathrm{10^7\, cm^2\, s^{-1}}$ to match the conditions used in \cite{Zahnle2016}. The atmosphere of 51 Eri b in both models is treated as having solar metallicity and a constant gravity of $32\, \mathrm{m\, s^{-2}}$. Since the temperature of 51 Eri b falls below $500\, \mathrm{K}$, the lower limit for convergence of our model, \textsc{Levi}, we solve for this atmosphere using \textsc{Argo} (\citealt{Rimmer2016}). 
 
Figure \ref{fig:51Moses} shows a close match between our model and that of \cite{Moses2016} for the carbon species, though we do find significantly less \ce{NH3} and significantly more \ce{OH}. Our comparison with \cite{Zahnle2016} in Figure \ref{fig:51Zahnle} shows large differences of at least an order of magnitude in the abundance of all of the sulfur species. Additionally, throughout much of the lower atmosphere, below $\mathrm{10^{-4}\,bar}$, we find nearly an order of magnitude more \ce{CO2} than \cite{Zahnle2016}. This negates most of the difference in abundance of \ce{CO2} that we initially expected to find from between sulfurous and non-sulfurous models in the deeper atmosphere when comparing \cite{Moses2016} and \cite{Zahnle2016}.

\begin{figure}
        \centering
        \includegraphics[width=0.475\textwidth]{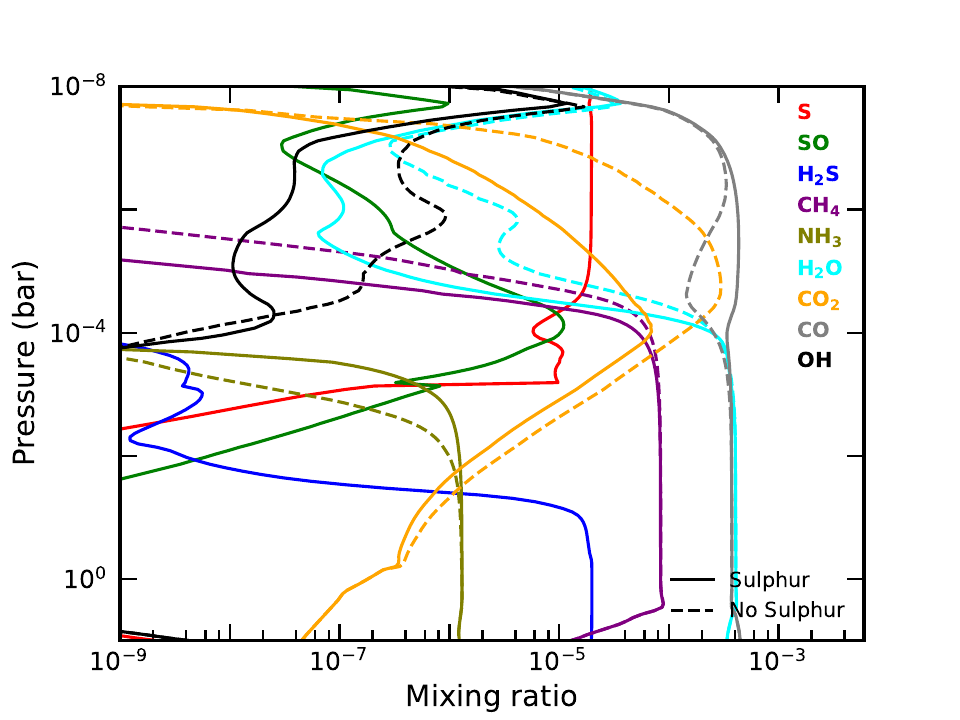}
    \caption[51EriSulfurComp]{A comparison of the abundance profiles for a warm Jupiter model both with (solid lines) and without (dashed lines) our new sulfur network being included. This warm Jupiter uses the P-T profile and UV spectrum from \cite{Moses2016}. It has a gravity of $32\, \mathrm{m\, s^{-2}}$, a solar metallicity and a constant $K_{zz} = \mathrm{10^7\, cm^2\, s^{-1}}$.}
     \label{fig:51EriSulfurComp}
\end{figure}
 
 However, in Figure \ref{fig:51EriSulfurComp} we compare two models, with and without sulfur, both using a constant $K_{zz} = \mathrm{10^7\, cm^2\, s^{-1}}$. Our results clearly show that while the \ce{CO2} abundance in both models below $\mathrm{10^{-4}\,bar}$ are similar, above this pressure, significant differences appear. In a sulfurous atmosphere \ce{CO2} is significantly depleted, with its peak abundance five times less than in a non-sulfurous atmosphere. This is less than the difference seen between \citet{Zahnle2016} and \citet{Moses2016}. Our results for \ce{CO_2} overall agree better with \citet{Moses2016}, both with and without sulfur. However, the reason for the decrease in \ce{CO2} abundance with the inclusion of sulfur is the effect of the products of \ce{H_2S} photo-dissociation on the abundance of \ce{OH}. Both with and without sulfur, \ce{CO_2} is produced by the following reactions
 
\begin{equation}
\begin{aligned}
\ce{H2O +}\ h\nu \ce{&$\, \rightarrow \, $ OH + H}, \\
\ce{CO} + \ce{OH} \ce{&$\, \rightarrow \, $} \ce{CO_2} + \ce{H}, \label{eqn:recombine}\\
\ce{H} + \ce{H} + \ce{M} \ce{&$\, \rightarrow \, $} \ce{H_2} + \ce{M}\\
\mathbf{Net:}\ \ce{H_2O} + \ce{CO} + h\nu \ce{&$\, \rightarrow \, $} \ce{CO_2} + \ce{H_2}.
\end{aligned}
\end{equation}

The destruction is also the same regardless of the presence of sulfur, through the photo-dissociation of \ce{CO_2}. What does change is the abundance of \ce{OH}, which is consumed by reacting with \ce{S} via

\begin{equation}
\begin{aligned}
\ce{H_2S} + h\nu \ce{&$\, \rightarrow \, $} \ce{HS} + \ce{H}, \\
\ce{HS} + \ce{H} \ce{&$\, \rightarrow \, $} \ce{H_2} + \ce{S}, \\
\ce{S} + \ce{OH} \ce{&$\, \rightarrow \, $} \ce{SO} + \ce{H}.
\end{aligned}
\end{equation}

The \ce{SO} will diffuse upward, where it will  photo-dissociate, releasing more atomic sulfur which will consume more hydroxyl radicals. It is this effect of the sulfur chemistry on the hydroxyl radicals that causes the lower abundance of \ce{CO_2} when sulfur is included. The destruction of \ce{OH} via the photochemical properties of \ce{H2S} explains the shift of the peak of \ce{CO_2} to higher pressures when sulfur is present.

There is one other feature of note, the \ce{H_2O} spike near $\mathrm{10^{-8}\,bar}$. This occurs because our constant $K_{zz}$ places the \ce{CO} homopause at $\sim 10^{-7} \, {\rm bar}$. As the \ce{CO} drops off, it stops effectively self-shielding, and so \ce{CO} is easily dissociated into \ce{C(^1D)} and \ce{O(^1D)}. The \ce{O(^1D)} reacts with \ce{H_2} to form \ce{H_2O}. \ce{H_2O} is preserved at this height, because its destruction forms \ce{OH} which quickly reacts with \ce{H2} to reform \ce{H2O}

\begin{equation}
\begin{aligned}
\ce{H_2} + \ce{OH} \ce{&$\, \rightarrow \, $} \ce{H_2O} + \ce{H}.
\end{aligned}
\end{equation}

This reaction dominates instead of the reaction \ce{CO + OH $\, \rightarrow \, $ CO2 + H} (See Reaction \ref{eqn:recombine}) since the depletion of \ce{CO} due to molecular diffusion means that it is no longer as efficient. This leads to \ce{H_2O} peaking at $\sim 10^{-8} \, {\rm bar}$. It drops off above this height because of a combination of molecular diffusion of \ce{H_2O} and photochemical destruction of \ce{OH}.
\section{Summary and Discussion} 

\label{Section6} 

Our goal in this work was to construct a complete sulfur chemical network that would primarily function for the high temperatures of hot Jupiters, but that should be applicable over a large range of temperatures. By combining this network with our chemical kinetics code, \textsc{Levi}, described in \cite{Hobbs2019} and the H/C/O/N network STAND2019  from \cite{Rimmer2016} and \cite{Rimmer2019} we gained the ability to find the steady-state solutions for chemical abundance in the atmospheres of a variety of hot Jupiters. We validated our new network against similar previously published models to identify the similarities and differences. We investigated the primary pathways of sulfur chemistry over a range of hot Jupiter models, and examined how the abundance of sulfur species changed with variations to the model. We also compared how the inclusion of sulfur into a hot Jupiter's atmosphere can affect the abundance of other, non-sulfur species, potentially altering these species detectability.

We started by validating our network against previous models, including the work of \cite{Wang2017} and the work presented in \cite{Evans2018} (but made by \citealt{Zahnle2016}). We found our results matched nearly perfectly with \cite{Wang2017} (Figure \ref{fig:Wang17Comp}), but their work only explored pressures below $\mathrm{10^{-4}\,bar}$, and so photochemistry was not considered. Similarly, we find that our results matched well with \cite{Evans2018} below $\mathrm{10^{-4}\,bar}$ (Figure \ref{fig:ZahnleWaspComp}). However, in the upper atmosphere, we find the species' abundances follow the same general trends, but at significantly different pressures. This is inferred to be most likely due to different choices in the UV flux applied to the model.

Across a range of isothermal hot Jupiter atmospheres, we investigated the most abundant sulfur species and how this varies with the model's parameters. In our disequilibrium atmosphere model (Figure \ref{fig:TripleDEQComp}) we found that from the bottom of the atmosphere until between $\mathrm{10^{-2}}$ and $\mathrm{10^{-3}\,bar}$, \ce{H2S} contains nearly all of the atmosphere's sulfur. Above this, \ce{H2S} is destroyed by photochemically produced \ce{H} radicals that cause atomic \ce{S} to become the primary sulfur carrier for the rest of the atmosphere. Other sulfur molecules of note are \ce{S2} and \ce{HS}. \ce{S2} builds up significantly around $\mathrm{10^{-3}\,bar}$, particularly on slightly cooler hot Jupiters, where for a brief pressure window it is the most abundant sulfur species. The mercapto radical, \ce{HS}, is one of the few sulfur molecules whose presence may have already been detected on an exoplanet (\citealt{Evans2018}). We find that it builds up strongly around $\mathrm{10^{-3}\,bar}$, especially along isotherms $>1200\,\mathrm{K}$, with a maximum abundance of $\mathrm{4x10^{-3}}$ seen in our model in Figure \ref{fig:TripleDEQComp}. 

We also examined the sensitivity of sulfur chemistry to variations in both the strength of the vertical mixing (Figure \ref{fig:KzzSens}) and the strength of the UV flux (Figure \ref{fig:UVSens}) applied to the atmosphere. We found that weakening the diffusion had little effect on the sulfur species. Stronger mixing increased the abundance of \ce{H2S} above $\mathrm{10^{-3}\,bar}$, while decreasing the abundance of \ce{S} and \ce{S2}. The effect of stronger UV irradiation is primarily to dissociate photosensitive molecules deeper into the atmosphere. Species such as \ce{H2S} were up to two orders of magnitude less abundant at $\mathrm{10^{-3}\,bar}$ in more highly irradiated models.

Most sulfur species are not easily detected in exoplanet atmospheres, but that does not mean that sulfur chemistry does not matter for atmospheric retrievals. Our models have shown that the presence of sulfur in both hot and warm Jupiter atmospheres can significantly change the abundance of other, more easily detected, molecules. Of particular note are the differences we have found to the abundances of \ce{CH4}, \ce{HCN} and \ce{NH3} in hot Jupiters by including \ce{S} in the network, as well as \ce{CO2} in warm Jupiters. We found this was due to the catalytic properties of sulfur chemistry (also seen in \citealt{Zahnle2016}) for \ce{CH4} and \ce{HCN}, and due to the photo-shielding effect of \ce{H2S} for \ce{NH3}. At around a pressure of $\mathrm{10^{-3}\,bar}$ in a atmospheric model of HD 209458b (Figure \ref{fig:HD209SulfurComp}), the abundances of \ce{NH3}, \ce{CH4} and \ce{HCN} in a sulfurous atmosphere dropped by several orders of magnitude when compared to their abundances in a non-sulfurous atmosphere.  This is a very large change, but it occurs over a very narrow pressure window. If this intersects with the detectable range made by observations, it would be a significant and noticeable effect. But, if the detection occurred at location higher or lower in the atmosphere, no change would be noticed compared to a sulfur free case. It is also very likely that the location at which sulfur chemistry impacts these molecules is dependant upon many of the planet's parameters: The planet's temperature and metallicity, as well as the diffusion strength or the incoming UV flux. More testing and modelling is required to discover exactly how these parameters will alter sulfur's effect upon these molecules. 

The effect of sulfur upon \ce{CO2} in warm Jupiters is also important, first noticed when comparing the sulfurous (\citealt{Zahnle2016}) and non-sulfurous (\citealt{Moses2016}) models of 51 Eri b. We find our network predicts up to five times less \ce{CO2}, due to sulfur, above $\mathrm{10^{-4}\,bar}$. Below this pressure, the loss of \ce{CO2} is present, but less significant. In this section of the upper atmosphere, we predict approximately half an order of magnitude more \ce{CO2} compared to the model in \cite{Zahnle2016}. These differences in the models are again down to uncertainties in the rate constants used for our sulfur reactions.

Current observations may struggle to distinguish the effects of sulfur we predict in both hot and warm Jupiters. However, it is important that observations from the next generation of telescopes take these effects into account, since they can significantly alter the composition of the atmosphere. An understanding of the precise effects of sulfur can be difficult to reach, due to how many uncertainties there are in the reaction rate constants within the network. As we have shown, these rates can greatly affect not just other sulfur species, but other detectable species within the atmosphere. New measurements of the rate constants of these sulfur reactions need to take place to truly be able to constrain the exact effect that sulfur can have. Some of the most important reactions discussed within this work still have large divergences between their different measurements. We discuss where some of these reactions' rate constants come from in Appendix \ref{app:ratesdiscuss}.

Finally, it is important to consider that while it is difficult to identify the bulk sulfur abundance without the detection of sulfur species, the effect of sulfur upon other species can be used as a method to constrain the sulfur abundance. There of course would be a large degree of degeneracy in this parameter, but it could work as a first step to limiting the sulfur abundance of exo-planets to realistic values.

\section*{Acknowledgements}

R.H. and O.S. acknowledge support from the UK Science and Technology Facilities Council (STFC). P.B.R. thanks the Simons Foundation for funding (SCOL awards 599634)

\section*{Data availability}
The data underlying this article is available in the article.

%%%%%%%%%%%%%%%%%%%%%%%%%%%%%%%%%%%%%%%%%%%%%%%%%%



\bibliographystyle{mnras}
\bibliography{bibliography,reactions}

\appendix
\section{Rates Discussion}
\label{app:ratesdiscuss}

Here we discuss some of the most significant sulfur chemical reaction rates in our network, their uncertainties, and where these uncertainties come from.

{\bf Photochemistry of \ce{H_2S}}\\
The photo-dissociation of \ce{H2S} is one of two photochemical sulfur reactions that form the explanation for sulfur's significant impact on other species in warm and hot Jupiters. There are a large number of works that have measured its UV cross-section, though not all of them agree what that cross-section should be. 
The work of \citet{Lee1987} provides an effective ``backbone'' to the wavelength-dependent absorption cross-sections for \ce{H_2S}, and provides the lowest measured estimates for \ce{H_2S} absorption of all observations, diverging from other measurements at $\lambda < 172$ nm. Both \citet{Watanabe1964} and \citet{Wu1998} find cross-sections below 172 nm that are a factor of 2-3 greater. 
\citet{Thompson1966} measures the absorption cross-section between 180 nm and 214 nm, in good agreement with other measurements at these wavelengths, though with a slight divergence between 210 -- 214 nm. 
\citet{Wu1998} generally agrees with most other measurements wherever one of the measurements diverges, and matches a measurement at a single (250 nm) wavelength very well \citep{Wight1983}. The measurements of \citet{Wu1998} are clean and the results suggest few systematics and a very deep noise floor. For this reason the JPL database recommends using their \ce{H_2S} cross-sections. 
More recently, \citet{Grosch2015} has made measurements at a higher resolution. These measurements diverge from \citet{Wu1998} by a factor of 5 between 250 and 260 nm, suggesting the higher resolution has come at the cost of a higher noise floor.

{\bf Photo-chemistry of \ce{SO}}\\
The photo-dissociation of \ce{SO} is the second of the important photochemical sulfur reactions. The most accurate cross-sections for \ce{SO} were compiled by the Leiden database \citep{Heays2017}, based on \citet{Phillips1981,Nee1986,Norwood1989}. The cross-sections all agree with each other reasonably well.

{\bf H + SH -> \ce{H2} + S} vs {\bf \ce{H2} + S -> H + SH}\\
We found that the reaction \ce{H + SH -> H2 + S} was significant in the pathway that lead to \ce{CO2} destruction in 51 Eri b's atmosphere. However, in our network we choose to use this reaction's reverse, \ce{H2 + S -> H + SH}, even though it has an activation barrier. Our justification for this is that for \ce{H + SH -> H2 + S} all the results are at room temperature, and all of them were measured in the 1970's and 1980's.

\citet{Cupitt1970} performed a flow discharge with \ce{H_2S} and \ce{O_2}, and estimated the rate to be $k = 1.3 \times 10^{-10} \, {\rm cm^3 \, s^{-1}}$. The experiment was subsequently revisited, and the rate constant corrected to $k = 2.51 \times 10^{-11} \, {\rm cm^3 \, s^{-1}}$ in \citep{Cupitt1975}.

\citet{Bradley1973} also performed a flow discharge with \ce{H_2S}, controlling the overall reaction with \ce{NO}, attempting to isolate the individual reactions, and constrained the \ce{HS} hydrogen abstraction reaction to $k = 4.15 \times 10^{-11} \, {\rm cm^3 \, s^{-1}}$. Their results do not indicate any kinetic barrier or polynomial temperature dependence.

\citet{Tiee1981} used a 193 nm UV laser to dissociate \ce{H_2S}, forming \ce{HS}, and then subsequently measured the decay of \ce{HS}. They found a maximum rate of $k < 1.69 \times 10^{-11} \, {\rm cm^3 \, s^{-1}}$ .

\citet{Nicholas1979} used radio-frequency pulse discharge and found the measurement of \ce{HS} decay to be $k = 2.16 \times 10^{-11} \, {\rm cm^3 \, s^{-1}}$.

\citet{Langford1972} used flash photolysis and made the measurement of the \ce{HS} decay to be $k = 1.1 \times 10^{-11} \, {\rm cm^3 \, s^{-1}}$.

The reported rates are highly discordant, nor do they address the temperature dependence. In contrast, the reverse reaction, \ce{H2 + S -> H + SH}, was constrained much better by two experiments in the late 1990's. They were performed over a much larger temperature range, one more applicable to hot Jupiters.

\citet{Woiki1995} used \ce{OCS}/\ce{H2} pyrolysis and \ce{CS2}/\ce{H2} photolysis to generate S atoms that could then react with \ce{H2}. They found the reaction rate to be $k = 9.96 \times 10^{-10} \exp(-12070\mathrm{K}/\mathrm{T}) \, {\rm cm^3 \, s^{-1}}$ for a temperature range of $1257\, \mathrm{K}$ to $3137\, \mathrm{K}$.

\citet{Shiina1996} studied the thermal decomposition of \ce{H2S} and its subsequent reactions, and found the reaction between \ce{H2} and \ce{S} to be $k = 2.62 \times 10^{-10} \exp(-9920\mathrm{K}/\mathrm{T}) \, {\rm cm^3 \, s^{-1}}$ over a temperature range of $1050\, \mathrm{K}$ to $1660\, \mathrm{K}$. 

As a result we decided that using the reaction \ce{H2 + S -> H + SH} in our network was the choice offering the greatest accuracy in modelling hot Jupiters.

{\bf OH + S -> H + SO}\\
This reaction is the last reaction important to explaining the impact of sulfur in hot and warm Jupiters. It has been mentioned in several reviews, all based on a single experimental measurement by \citet{Jourdain1979}. They used a flow discharge with \ce{SO_2} and \ce{H_2O}, and found $k = 6.95 \times 10^{-11} \, {\rm cm^3 \, s^{-1}}$. Without a second measurement, it is difficult to determine this reaction's uncertainties, and without uncertainties, there is no way to perform a sensitivity analysis accurately. 

\section{The Sulfur Network}
\label{The sulfur network}

\begin{table*}
\centering
\caption{The 3-body reactions of the sulphur network used in this work. $\alpha$, $\beta$ and $\gamma \, (\mathrm{K})$ refer to the three constants used in the arrhenius equation, such that the rate of reaction is $k=\alpha (T/300\, \mathrm{K})^{\beta} \exp(-\gamma/T)\, \mathrm{m^3 s^{-1}}$. The reactions in this table are three body combination reactions that use a third body (M) as a catalyst. The first line of each reaction is the low pressure rate limit ($k_0$) and the second line is the high pressure rate limit ($k_\infty$) that are combined to give an overall rate via $k=\frac{k_0 [M]}{1+\frac{k_0[M]}{K_\infty}}$. Thus $\alpha_0$ has units $\mathrm{m^6 s^{-1}}$ and $\alpha_\infty$ has units $\mathrm{m^3 s^{-1}}$. }
\label{tab:network2}
\begin{tabular}{lcccccccccrrrcl}
\hline
No. & &&& Reaction &&&&&& \multicolumn{1}{c}{$\alpha\, (m^6 s^{-1}\, \textrm{or}\, m^3 s^{-1})$}&\multicolumn{1}{c}{$\beta$}&\multicolumn{1}{c}{$\gamma \, (\mathrm{K})$}& Reference\\
\hline
1  &  \ce{S} &+& \ce{S} &$\rightarrow$& \ce{S2} &&&&&3.95$\times10^{-45} $&&  &\cite{Du2008}  \\     
   &  \ce{S} &+& \ce{S} &$\rightarrow$& \ce{S2} &&&&&9.09$\times10^{-20} $&&  &\cite{Du2008}  \\        
2  &  \ce{S} &+& \ce{S2} &$\rightarrow$& \ce{S3} &&&&&1.11$\times10^{-42} $&-2.00&  &\cite{Moses2002}  \\     
   &  \ce{S} &+& \ce{S2} &$\rightarrow$& \ce{S3} &&&&&3$\times10^{-17} $&&  &\cite{Moses2002}  \\     
3  &  \ce{S} &+& \ce{S3} &$\rightarrow$& \ce{S4} &&&&&1.11$\times10^{-42} $&-2.00&  &\cite{Moses2002}  \\     
   &  \ce{S} &+& \ce{S3} &$\rightarrow$& \ce{S4} &&&&&3$\times10^{-17} $&&  &\cite{Moses2002}  \\   
4  &  \ce{S} &+& \ce{S4} &$\rightarrow$& \ce{S5} &&&&&1.11$\times10^{-42} $&-2.00&  &\cite{Moses2002}  \\     
   &  \ce{S} &+& \ce{S4} &$\rightarrow$& \ce{S5} &&&&&3$\times10^{-17} $&&  &\cite{Moses2002}  \\         
5  &  \ce{S} &+& \ce{S5} &$\rightarrow$& \ce{S6} &&&&&1.11$\times10^{-42} $&-2.00&  &\cite{Moses2002}  \\     
   &  \ce{S} &+& \ce{S5} &$\rightarrow$& \ce{S6} &&&&&3$\times10^{-17} $&&  &\cite{Moses2002}  \\     
6  &  \ce{S} &+& \ce{S6} &$\rightarrow$& \ce{S7} &&&&&1.11$\times10^{-42} $&-2.00&  &\cite{Moses2002}  \\     
   &  \ce{S} &+& \ce{S6} &$\rightarrow$& \ce{S7} &&&&&3$\times10^{-17} $&&  &\cite{Moses2002}  \\     
7  &  \ce{S} &+& \ce{S7} &$\rightarrow$& \ce{S8} &&&&&1.11$\times10^{-42} $&-2.00&  &\cite{Moses2002}  \\     
   &  \ce{S} &+& \ce{S7} &$\rightarrow$& \ce{S8} &&&&&3$\times10^{-17} $&&  &\cite{Moses2002}  \\     
8  &  \ce{S2} &+& \ce{S2} &$\rightarrow$& \ce{S4} &&&&&2.2$\times10^{-41} $&&  &\cite{Nicholas1979}  \\     
   &  \ce{S2} &+& \ce{S2} &$\rightarrow$& \ce{S4} &&&&&1$\times10^{-16} $&&  &\cite{Nicholas1979}  \\     
9  &  \ce{S2} &+& \ce{S3} &$\rightarrow$& \ce{S5} &&&&&1.11$\times10^{-42} $&-2.00&  &\cite{Moses2002}  \\     
   &  \ce{S2} &+& \ce{S3} &$\rightarrow$& \ce{S5} &&&&&3$\times10^{-17} $&&  &\cite{Moses2002}  \\     
10 &  \ce{S2} &+& \ce{S4} &$\rightarrow$& \ce{S6} &&&&&1.11$\times10^{-42} $&-2.00&  &\cite{Moses2002}  \\     
   &  \ce{S2} &+& \ce{S4} &$\rightarrow$& \ce{S6} &&&&&3$\times10^{-17} $&&  &\cite{Moses2002}  \\     
11 &  \ce{S2} &+& \ce{S5} &$\rightarrow$& \ce{S7} &&&&&1.11$\times10^{-42} $&-2.00&  &\cite{Moses2002}  \\     
   &  \ce{S2} &+& \ce{S5} &$\rightarrow$& \ce{S7} &&&&&3$\times10^{-17} $&&  &\cite{Moses2002}  \\     
12 &  \ce{S2} &+& \ce{S6} &$\rightarrow$& \ce{S8} &&&&&1.11$\times10^{-42} $&-2.00&  &\cite{Moses2002}  \\     
   &  \ce{S2} &+& \ce{S6} &$\rightarrow$& \ce{S8} &&&&&3$\times10^{-17} $&&  &\cite{Moses2002}  \\     
13 &  \ce{S3} &+& \ce{S3} &$\rightarrow$& \ce{S6} &&&&&1$\times10^{-42} $&&  &\cite{Mills1998}  \\     
   &  \ce{S3} &+& \ce{S3} &$\rightarrow$& \ce{S6} &&&&&3$\times10^{-17} $&&  &\cite{Mills1998}  \\     
14 &  \ce{S3} &+& \ce{S4} &$\rightarrow$& \ce{S7} &&&&&1.11$\times10^{-42} $&-2.00&  &\cite{Moses2002}  \\     
   &  \ce{S3} &+& \ce{S4} &$\rightarrow$& \ce{S7} &&&&&3$\times10^{-17} $&&  &\cite{Moses2002}  \\     
15 &  \ce{S3} &+& \ce{S5} &$\rightarrow$& \ce{S8} &&&&&1.11$\times10^{-42} $&-2.00&  &\cite{Moses2002}  \\     
   &  \ce{S3} &+& \ce{S5} &$\rightarrow$& \ce{S8} &&&&&3$\times10^{-17} $&&  &\cite{Moses2002}  \\     
16 &  \ce{S4} &+& \ce{S4} &$\rightarrow$& \ce{S8} &&&&&1$\times10^{-42} $&&  &\cite{Mills1998}  \\     
   &  \ce{S4} &+& \ce{S4} &$\rightarrow$& \ce{S8} &&&&&3$\times10^{-17} $&&  &\cite{Mills1998}  \\     
17 &  \ce{S} &+& \ce{O} &$\rightarrow$& \ce{SO} &&&&&3.01$\times10^{-45} $&&  &\cite{Zhang2012}  \\     
   &  \ce{S} &+& \ce{O} &$\rightarrow$& \ce{SO} &&&&&7.27$\times10^{-20} $&-1.00&  &\cite{Zhang2012}  \\
18 &  \ce{S} &+& \ce{SO} &$\rightarrow$& \ce{S2O} &&&&&3.67$\times10^{-43} $&-2.00&  &\cite{Moses2002}  \\     
   &  \ce{S} &+& \ce{SO} &$\rightarrow$& \ce{S2O} &&&&&8.86$\times10^{-19} $&-3.00&  &\cite{Moses2002}  \\ 
19 &  \ce{S} &+& \ce{H2} &$\rightarrow$& \ce{H2S} &&&&&1.4$\times10^{-43} $&-1.9&2.3$\times10^{3}$  &\cite{Zahnle2016}  \\     
   &  \ce{S} &+& \ce{H2} &$\rightarrow$& \ce{H2S} &&&&&1$\times10^{-17} $& &  &\cite{Zahnle2016}  \\ 
20 &  \ce{SO} &+& \ce{HO} &$\rightarrow$& \ce{HOSO} &&&&&6.45$\times10^{-41} $&-3.48&4.90$\times10^{2} $  &\cite{Goumri1999}   \\     
   &  \ce{SO} &+& \ce{HO} &$\rightarrow$& \ce{HOSO} &&&&&8.75$\times10^{-17} $&0.50&  &\cite{Goumri1999}  \\     
21 &  \ce{SO2} &+& \ce{O} &$\rightarrow$& \ce{SO3} &&&&&1.32$\times10^{-41} $&&1.00$\times10^{3} $  &\cite{Atkinson2004}   \\     
   &  \ce{SO2} &+& \ce{O} &$\rightarrow$& \ce{SO3} &&&&&5$\times10^{-18} $&&  &\cite{Atkinson2004}   \\     
22 &  \ce{SO2} &+& \ce{H} &$\rightarrow$& \ce{HSO2} &&&&&5.74$\times10^{-43} $&-3.69&2.41$\times10^{3} $  &\cite{Goumri1999}     \\     
   &  \ce{SO2} &+& \ce{H} &$\rightarrow$& \ce{HSO2} &&&&&2.31$\times10^{-17} $&0.62&1.82$\times10^{3} $  &\cite{Goumri1999}    \\     
23 &  \ce{SO2} &+& \ce{H} &$\rightarrow$& \ce{HOSO} &&&&&9.43$\times10^{-40} $&-4.36&5.44$\times10^{3} $  &\cite{Goumri1999}   \\     
   &  \ce{SO2} &+& \ce{H} &$\rightarrow$& \ce{HOSO} &&&&&9.13$\times10^{-18} $&0.96&4.32$\times10^{3} $  &\cite{Goumri1999}   \\     
24 &  \ce{SO} &+& \ce{O} &$\rightarrow$& \ce{SO2} &&&&&4.82$\times10^{-43} $&-2.17&  &\cite{Lu2003}  \\     
   &  \ce{SO} &+& \ce{O} &$\rightarrow$& \ce{SO2} &&&&&3.5$\times10^{-17} $&0.00&  &\cite{Lu2003}  \\     
25 &  \ce{SO2} &+& \ce{HO} &$\rightarrow$& \ce{HSO3} &&&&&3.3$\times10^{-43} $&-4.30&  &\cite{Sanders2006}   \\     
   &  \ce{SO2} &+& \ce{HO} &$\rightarrow$& \ce{HSO3} &&&&&1.6$\times10^{-18} $&0.00&  &\cite{Sanders2006}    \\     
26 &  \ce{HS} &+& \ce{NO} &$\rightarrow$& \ce{HSNO} &&&&&2.4$\times10^{-43} $&-3.00&  &\cite{Demore1997}   \\     
   &  \ce{HS} &+& \ce{NO} &$\rightarrow$& \ce{HSNO} &&&&&2.71$\times10^{-17} $&&  &\cite{Demore1997}   \\     
27 &  \ce{HS} &+& \ce{O2} &$\rightarrow$& \ce{HSO2} &&&&&9.18$\times10^{-46} $&-1.69&  &\cite{Goumri1995}  \\     
   &  \ce{HS} &+& \ce{O2} &$\rightarrow$& \ce{HSO2} &&&&&2.01$\times10^{-16} $&0.31&  &\cite{Goumri1995} \\     
28 &  \ce{HS} &+& \ce{O2} &$\rightarrow$& \ce{HSOO} &&&&&9.06$\times10^{-46} $&-2.01&1.00$\times10^{1} $  &\cite{Goumri1999}  \\     
   &  \ce{HS} &+& \ce{O2} &$\rightarrow$& \ce{HSOO} &&&&&3.3$\times10^{-16} $&-0.26&1.50$\times10^{2} $  &\cite{Goumri1999}  \\     
29 &  \ce{HS} &+& \ce{H} &$\rightarrow$& \ce{H2S} &&&&&1$\times10^{-42} $&-2.00&  &\cite{Krasnopolsky2007}  \\     
   &  \ce{HS} &+& \ce{H} &$\rightarrow$& \ce{H2S} &&&&&2.41$\times10^{-17} $&-3.00&  &\cite{Krasnopolsky2007}  \\     
30 &  \ce{CS2} &+& \ce{HO} &$\rightarrow$& \ce{CS2OH} &&&&&8$\times10^{-43} $&&  &\cite{Atkinson1992}  \\     
   &  \ce{CS2} &+& \ce{HO} &$\rightarrow$& \ce{CS2OH} &&&&&8$\times10^{-18} $&&  &\cite{Atkinson1992}  \\     
31 &  \ce{CO} &+& \ce{S} &$\rightarrow$& \ce{OCS} &&&&&3$\times10^{-45} $&&1.00$\times10^{3} $  &\cite{Krasnopolsky2007}  \\     
   &  \ce{CO} &+& \ce{S} &$\rightarrow$& \ce{OCS} &&&&&7.24$\times10^{-20} $&-1.00&1.00$\times10^{3} $  &\cite{Krasnopolsky2007}  \\     
32 &  \ce{CH3} &+& \ce{HS} &$\rightarrow$& \ce{CH3SH} &&&&&6.88$\times10^{-43} $&1.00&  &\cite{Shum1985}   \\     
   &  \ce{CH3} &+& \ce{HS} &$\rightarrow$& \ce{CH3SH} &&&&&1.66$\times10^{-17} $&&  &\cite{Shum1985}   \\   
   
   \hline
\end{tabular}
\end{table*} 

\begin{table*}
\centering
\caption{The 3-body reactions of the sulphur network used in this work. $\alpha$, $\beta$ and $\gamma \, (\mathrm{K})$ refer to the three constants used in the arrhenius equation, such that the rate of reaction is $k=\alpha (T/300\, \mathrm{K})^{\beta} \exp(-\gamma/T)\, \mathrm{m^3 s^{-1}}$. The reactions in this table are three body decomposition reactions that use a third body (M) as a catalyst and a state shift reaction. The first line of each reaction is the low pressure rate limit ($k_0$) and the second line is the high pressure rate limit ($k_\infty$) that are combined to give an overall rate via $k=\frac{k_0 [M]}{1+\frac{k_0[M]}{K_\infty}}$. Thus $\alpha_0$ has units $\mathrm{m^3 s^{-1}}$ and $\alpha_\infty$ has units $\mathrm{s^{-1}}$.}
\label{tab:network3}
\begin{tabular}{lcccccccccrrrcl}
\hline
No. & &&& Reaction &&&&&&\multicolumn{1}{c}{$\alpha\, (m^3 s^{-1}\, \textrm{or}\, s^{-1})$}&\multicolumn{1}{c}{$\beta$}&\multicolumn{1}{c}{$\gamma \, (\mathrm{K})$}& Reference\\
\hline

33 &  \ce{HSO} &&&$\rightarrow$& \ce{H} &+& \ce{SO} &&&1.4$\times10^{-14} $&&2.95$\times10^{4} $  &\cite{Tsuchiya1994}   \\     
   &  \ce{HSO} &&&$\rightarrow$& \ce{H} &+& \ce{SO} &&&3.38$\times10^{+11} $&-1.00&2.95$\times10^{4} $  &\cite{Tsuchiya1994}   \\     
34 &  \ce{HOSO} &&&$\rightarrow$& \ce{HSO2} &&&&&3.18$\times10^{-09} $&-5.64&2.79$\times10^{4} $  &\cite{Goumri1999}   \\     
   &  \ce{HOSO} &&&$\rightarrow$& \ce{HSO2} &&&&&3.64$\times10^{+11} $&1.03&2.52$\times10^{4} $  &\cite{Goumri1999}  \\     
35 &  \ce{HSOO} &&&$\rightarrow$& \ce{O} &+& \ce{HSO} &&&4.61$\times10^{-10} $&-5.87&1.56$\times10^{4} $  &\cite{Goumri1999}   \\     
   &  \ce{HSOO} &&&$\rightarrow$& \ce{O} &+& \ce{HSO} &&&4.53$\times10^{+16} $&-1.07&1.43$\times10^{4} $  &\cite{Goumri1999}  \\     
36 &  \ce{H2S2} &&&$\rightarrow$& \ce{HS} &+& \ce{HS} &&&3.43$\times10^{-13} $&1.00&2.87$\times10^{4} $  &\cite{Tsuchiya1994}   \\     
   &  \ce{H2S2} &&&$\rightarrow$& \ce{HS} &+& \ce{HS} &&&8.28$\times10^{+12} $&&2.87$\times10^{4} $  &\cite{Tsuchiya1994}  \\  
   \hline
\end{tabular}
\end{table*} 

\begin{table}
\centering
\caption{The photo-chemical reactions of the sulphur network used in this work.}
\label{tab:network1}
\begin{tabular}{lccccc}
\hline
No. & & Reaction &&&  \\
\hline 
   
1   &  \ce{S3}    &$\rightarrow$& \ce{S2} &+& \ce{S}               \\     
2   &  \ce{S4}    &$\rightarrow$& \ce{S2} &+& \ce{S2}              \\     
3   &  \ce{SO3}   &$\rightarrow$& \ce{SO2} &+& \ce{O}             \\     
4   &  \ce{OCS}   &$\rightarrow$& \ce{S} &+& \ce{CO}              \\     
5   &  \ce{OCS}   &$\rightarrow$& \ce{S} &+& \ce{CO}              \\     
6   &  \ce{OCS}   &$\rightarrow$& \ce{S} &+& \ce{CO}              \\     
7   &  \ce{OCS}   &$\rightarrow$& \ce{CS} &+& \ce{O}           \\     
8   &  \ce{OCS}   &$\rightarrow$& \ce{CS} &+& \ce{O(^1D)}        \\     
9   &  \ce{S2}    &$\rightarrow$& \ce{S} &+& \ce{S}                \\     
10   &  \ce{S2O}   &$\rightarrow$& \ce{SO} &+& \ce{S}              \\     
11  &  \ce{S2O}   &$\rightarrow$& \ce{S2} &+& \ce{O}           \\     
12   &  \ce{SO}    &$\rightarrow$& \ce{S} &+& \ce{O}                \\     
13   &  \ce{CS2}   &$\rightarrow$& \ce{CS} &+& \ce{S}              \\     
14   &  \ce{SO2}   &$\rightarrow$& \ce{SO} &+& \ce{O}              \\     
15   &  \ce{SO2}   &$\rightarrow$& \ce{S} &+& \ce{O2}              \\     
16   &  \ce{H2S}   &$\rightarrow$& \ce{HS} &+& \ce{H}              \\     
17   &  \ce{CH3SH} &$\rightarrow$& \ce{CH3} &+& \ce{HS}       \\                  
\hline
\end{tabular}
\end{table}

\begin{table*}
\centering
\caption{The 2-body reactions of the sulphur network used in this work. $\alpha\,(\mathrm{m^3 s^{-1}})$, $\beta$ and $\gamma \, (\mathrm{K})$ refer to the three constants used in the arrhenius equation, such that the rate of reaction is $k=\alpha (T/300\, \mathrm{K})^{\beta} \exp(-\gamma/T)\, \mathrm{m^3 s^{-1}}$. For the reactions with multiple sources, we have blended data from both sources to extend the applicable temperature range for the reaction.}
\label{tab:network4}
\begin{tabular}{lcccccccccrrrcl}
\hline
No. & &&& Reaction &&&&&&\multicolumn{1}{c}{$\alpha\, (m^3 s^{-1})$}&\multicolumn{1}{c}{$\beta$}&\multicolumn{1}{c}{$\gamma \, (\mathrm{K})$}& Reference\\
\hline
1    &  \ce{S} &+& \ce{S3} &$\rightarrow$& \ce{S2} &+& \ce{S2} &&&8$\times10^{-17} $&&  &\cite{Moses2002}  \\ 
2    &  \ce{S} &+& \ce{S4} &$\rightarrow$& \ce{S2} &+& \ce{S3} &&&8$\times10^{-17} $&&  &\cite{Moses2002}  \\ 
3    &  \ce{S} &+& \ce{S5} &$\rightarrow$& \ce{S2} &+& \ce{S4} &&&5$\times10^{-17} $&&2.00$\times10^{2} $  &\cite{Moses2002}  \\   
4    &  \ce{S} &+& \ce{S6} &$\rightarrow$& \ce{S2} &+& \ce{S5} &&&5$\times10^{-17} $&&3.00$\times10^{2} $  &\cite{Moses2002}  \\    
5    &  \ce{S} &+& \ce{S7} &$\rightarrow$& \ce{S2} &+& \ce{S6} &&&4$\times10^{-17} $&&2.00$\times10^{2} $  &\cite{Moses2002}  \\      
6    &  \ce{S} &+& \ce{S8} &$\rightarrow$& \ce{S2} &+& \ce{S7} &&&4$\times10^{-17} $&&4.00$\times10^{2} $  &\cite{Moses2002}  \\  
7    &  \ce{S} &+& \ce{S5} &$\rightarrow$& \ce{S3} &+& \ce{S3} &&&3$\times10^{-17} $&&2.00$\times10^{2} $  &\cite{Moses2002}  \\ 
8    &  \ce{S} &+& \ce{S6} &$\rightarrow$& \ce{S3} &+& \ce{S4} &&&3$\times10^{-17} $&&3.00$\times10^{2} $  &\cite{Moses2002}  \\     
9    &  \ce{S} &+& \ce{S7} &$\rightarrow$& \ce{S3} &+& \ce{S5} &&&2$\times10^{-17} $&&2.00$\times10^{2} $  &\cite{Moses2002}  \\     
10   &  \ce{S} &+& \ce{S8} &$\rightarrow$& \ce{S3} &+& \ce{S6} &&&2$\times10^{-17} $&&4.00$\times10^{2} $  &\cite{Moses2002}  \\     
11   &  \ce{S} &+& \ce{S7} &$\rightarrow$& \ce{S4} &+& \ce{S4} &&&2$\times10^{-17} $&&2.00$\times10^{2} $  &\cite{Moses2002}  \\       
12   &  \ce{S} &+& \ce{S8} &$\rightarrow$& \ce{S4} &+& \ce{S5} &&&2$\times10^{-17} $&&4.00$\times10^{2} $  &\cite{Moses2002}  \\     
13   &  \ce{S2} &+& \ce{S8} &$\rightarrow$& \ce{S5} &+& \ce{S5} &&&1$\times10^{-17} $&&1.40$\times10^{3} $  &\cite{Moses2002}  \\        
14   &  \ce{S3} &+& \ce{S5} &$\rightarrow$& \ce{S2} &+& \ce{S6} &&&4$\times10^{-17} $&&2.00$\times10^{2} $  &\cite{Moses2002}  \\     
15   &  \ce{S3} &+& \ce{S7} &$\rightarrow$& \ce{S2} &+& \ce{S8} &&&3$\times10^{-17} $&&2.00$\times10^{2} $  &\cite{Moses2002}  \\     
16   &  \ce{S3} &+& \ce{S7} &$\rightarrow$& \ce{S4} &+& \ce{S6} &&&1$\times10^{-17} $&&2.00$\times10^{2} $  &\cite{Moses2002}  \\     
17   &  \ce{S3} &+& \ce{S4} &$\rightarrow$& \ce{S2} &+& \ce{S5} &&&4$\times10^{-17} $&&2.00$\times10^{2} $  &\cite{Moses2002}  \\     
18   &  \ce{S3} &+& \ce{S6} &$\rightarrow$& \ce{S2} &+& \ce{S7} &&&4$\times10^{-18} $&&3.00$\times10^{2} $  &\cite{Moses2002}  \\     
19   &  \ce{S3} &+& \ce{S7} &$\rightarrow$& \ce{S5} &+& \ce{S5} &&&1$\times10^{-17} $&&2.00$\times10^{2} $  &\cite{Moses2002}  \\      
20   &  \ce{S4} &+& \ce{S5} &$\rightarrow$& \ce{S2} &+& \ce{S7} &&&2$\times10^{-18} $&&2.00$\times10^{2} $  &\cite{Moses2002}  \\     
21   &  \ce{S4} &+& \ce{S6} &$\rightarrow$& \ce{S2} &+& \ce{S8} &&&2$\times10^{-18} $&&3.00$\times10^{2} $  &\cite{Moses2002}  \\     
22   &  \ce{S4} &+& \ce{S5} &$\rightarrow$& \ce{S3} &+& \ce{S6} &&&2$\times10^{-18} $&&2.00$\times10^{2} $  &\cite{Moses2002}  \\     
23   &  \ce{S4} &+& \ce{S7} &$\rightarrow$& \ce{S3} &+& \ce{S8} &&&5$\times10^{-18} $&&2.00$\times10^{2} $  &\cite{Moses2002}  \\     
24   &  \ce{S4} &+& \ce{S6} &$\rightarrow$& \ce{S5} &+& \ce{S5} &&&2$\times10^{-18} $&&3.00$\times10^{2} $  &\cite{Moses2002}  \\     
25   &  \ce{S4} &+& \ce{S7} &$\rightarrow$& \ce{S5} &+& \ce{S6} &&&5$\times10^{-18} $&&2.00$\times10^{2} $  &\cite{Moses2002}  \\               
26   &  \ce{O} &+& \ce{S2} &$\rightarrow$& \ce{S}  &+& \ce{SO} &&&2$\times10^{-17} $&&8.40$\times10^{1} $  &\cite{Craven1987}  \\     
27   &  \ce{O} &+& \ce{S3} &$\rightarrow$& \ce{S2} &+& \ce{SO} &&&8$\times10^{-17} $&&  &\cite{Moses2002}  \\             
28   &  \ce{O} &+& \ce{S4} &$\rightarrow$& \ce{S3} &+& \ce{SO} &&&8$\times10^{-17} $&&  &\cite{Moses2002}  \\         
29   &  \ce{O} &+& \ce{S5} &$\rightarrow$& \ce{S4} &+& \ce{SO} &&&8$\times10^{-17} $&&2.00$\times10^{2} $  &\cite{Moses2002}  \\           
30   &  \ce{O} &+& \ce{S6} &$\rightarrow$& \ce{S5} &+& \ce{SO} &&&8$\times10^{-17} $&&3.00$\times10^{2} $  &\cite{Moses2002}  \\        
31   &  \ce{O} &+& \ce{S7} &$\rightarrow$& \ce{S6} &+& \ce{SO} &&&8$\times10^{-17} $&&2.00$\times10^{2} $  &\cite{Moses2002}  \\          
32   &  \ce{O} &+& \ce{S8} &$\rightarrow$& \ce{S7} &+& \ce{SO} &&&8$\times10^{-17} $&&4.00$\times10^{2} $  &\cite{Moses2002}  \\     
33   &  \ce{S} &+& \ce{O2} &$\rightarrow$& \ce{SO} &+& \ce{O} &&&2.51$\times10^{-17} $&&1.84$\times10^{3} $  &\cite{Miyoshi1996}  \\     
34   &  \ce{S} &+& \ce{SO2} &$\rightarrow$& \ce{SO} &+& \ce{SO} &&&9.77$\times10^{-18} $&&4.54$\times10^{3} $  &\cite{Isshiki2003}  \\     
35   &  \ce{S} &+& \ce{H2} &$\rightarrow$& \ce{H} &+& \ce{HS} &&&3.04$\times10^{-19} $&2.70&6.46$\times10^{3} $  &\cite{Shiina1996} and \cite{Woiki1995}  \\     
36   &  \ce{S} &+& \ce{OCS} &$\rightarrow$& \ce{CO} &+& \ce{S2} &&&1.35$\times10^{-19} $&2.70&1.20$\times10^{3} $  &\cite{Shiina1996} and \cite{Lu2006} \\     
37   &  \ce{S} &+& \ce{CS2} &$\rightarrow$& \ce{CS} &+& \ce{S2} &&&2.82$\times10^{-16} $&&5.92$\times10^{3} $  &\cite{Woiki1995}  \\     
38   &  \ce{S} &+& \ce{C2H6} &$\rightarrow$& \ce{HS} &+& \ce{C2H5} &&&2.04$\times10^{-16} $&&7.42$\times10^{3} $  &\cite{Tsuchiya1996}  \\     
39   &  \ce{S} &+& \ce{CH4} &$\rightarrow$& \ce{HS} &+& \ce{CH3} &&&3.39$\times10^{-16} $&&1.00$\times10^{4} $  &\cite{Tsuchiya1996}  \\     
40   &  \ce{S} &+& \ce{HS} &$\rightarrow$& \ce{S2} &+& \ce{H} &&&4.98$\times10^{-18} $&&  &\cite{Nicholas1979}  \\     
41   &  \ce{S} &+& \ce{HO2} &$\rightarrow$& \ce{SO} &+& \ce{HO} &&&5.84$\times10^{-17} $&&  &\cite{Zhang2012}  \\     
42   &  \ce{S} &+& \ce{SO3} &$\rightarrow$& \ce{SO2} &+& \ce{SO} &&&1$\times10^{-22} $&&  &\cite{Moses2002}  \\     
43   &  \ce{S} &+& \ce{HO} &$\rightarrow$& \ce{SO} &+& \ce{H} &&&6.6$\times10^{-17} $&&  &\cite{Demore1997}       \\     
44   &  \ce{S} &+& \ce{S2O} &$\rightarrow$& \ce{S2} &+& \ce{SO} &&&1$\times10^{-18} $&&1.20$\times10^{3} $  &\cite{Moses2002}  \\     
45   &  \ce{S} &+& \ce{O3} &$\rightarrow$& \ce{SO} &+& \ce{O2} &&&1.2$\times10^{-17} $&&  &\cite{Atkinson2004}  \\        
46   &  \ce{S3} &+& \ce{CO} &$\rightarrow$& \ce{OCS} &+& \ce{S2} &&&1$\times10^{-17} $&&2.00$\times10^{4} $  &\cite{Krasnopolsky2007}       \\     
47   &  \ce{S3} &+& \ce{H} &$\rightarrow$& \ce{HS} &+& \ce{S2} &&&1.2$\times10^{-16} $&&1.95$\times10^{3} $  &\cite{Krasnopolsky2007}       \\      
48   &  \ce{HS} &+& \ce{HS} &$\rightarrow$& \ce{H2S} &+& \ce{S} &&&1.5$\times10^{-17} $&&  &\cite{Schofield1973}         \\     
49   &  \ce{HS} &+& \ce{HO} &$\rightarrow$& \ce{H2O} &+& \ce{S} &&&2.5$\times10^{-18} $&&  &\cite{Krasnopolsky2007}       \\     
50   &  \ce{HS} &+& \ce{O} &$\rightarrow$& \ce{H} &+& \ce{SO} &&&1.3$\times10^{-16} $&&  &\cite{Tsuchiya1994}  \\     
51   &  \ce{HS} &+& \ce{NO2} &$\rightarrow$& \ce{NO} &+& \ce{HSO} &&&6.49$\times10^{-17} $&&  &\cite{Atkinson2004}  \\     
52   &  \ce{HS} &+& \ce{N2O} &$\rightarrow$& \ce{N2} &+& \ce{HSO} &&&5$\times10^{-22} $&&  &\cite{Herndon1993}  \\     
53   &  \ce{HS} &+& \ce{O3} &$\rightarrow$& \ce{O2} &+& \ce{HSO} &&&1.1$\times10^{-17} $&&2.80$\times10^{2} $  &\cite{Wang1990}   \\     
54   &  \ce{HS} &+& \ce{O2} &$\rightarrow$& \ce{O} &+& \ce{HSO} &&&3.11$\times10^{-17} $&&9.02$\times10^{3} $  &\cite{Tsuchiya1997}  \\     
55   &  \ce{HS} &+& \ce{O2} &$\rightarrow$& \ce{HO} &+& \ce{SO} &&&4$\times10^{-25} $&&  &\cite{Demore1997}  \\  
56   &  \ce{HS} &+& \ce{CO} &$\rightarrow$& \ce{OCS} &+& \ce{H} &&&4.15$\times10^{-20} $&&7.66$\times10^{3} $  &\cite{Kurbanov1995}  \\     
57   &  \ce{HS} &+& \ce{H2CS} &$\rightarrow$& \ce{H2S} &+& \ce{HCS} &&&8.14$\times10^{-17} $&&3.18$\times10^{3} $  &\cite{Vandeputte2010}  \\     
   \hline
\end{tabular}
\end{table*} 

\begin{table*}
\centering
\caption{The 2-body reactions of the sulphur network used in this work. $\alpha\, (\mathrm{m^3 s^{-1}})$, $\beta$ and $\gamma \, (\mathrm{K})$ refer to the three constants used in the arrhenius equation, such that the rate of reaction is $k=\alpha (T/300\, \mathrm{K})^{\beta} \exp(-\gamma/T)\, \mathrm{m^3 s^{-1}}$. }
\label{tab:network5}
\begin{tabular}{lcccccccccrrrcl}
\hline
No. & &&& Reaction &&&&&&\multicolumn{1}{c}{$\alpha\, (m^3 s^{-1})$}&\multicolumn{1}{c}{$\beta$}&\multicolumn{1}{c}{$\gamma \, (\mathrm{K})$}& Reference\\
\hline

58  &  \ce{H2S} &+& \ce{O} &$\rightarrow$& \ce{H} &+& \ce{HSO} &&&5$\times10^{-16} $&&3.85$\times10^{3} $  &\cite{Tsuchiya1994}  \\     
59   &  \ce{H2S} &+& \ce{O} &$\rightarrow$& \ce{HO} &+& \ce{HS} &&&2.01$\times10^{-16} $&&3.85$\times10^{3} $  &\cite{Tsuchiya1994}  \\     
60   &  \ce{H2S} &+& \ce{H} &$\rightarrow$& \ce{H2} &+& \ce{HS} &&&3.07$\times10^{-18} $&2.10&3.52$\times10^{2} $  &\cite{Yoshimura1992}  \\     
61   &  \ce{H2S} &+& \ce{O2} &$\rightarrow$& \ce{HO2} &+& \ce{HS} &&&3.1$\times10^{-18} $&2.76&1.92$\times10^{4} $  &\cite{Montoya2005}  \\     
62   &  \ce{H2S} &+& \ce{HO} &$\rightarrow$& \ce{H2O} &+& \ce{HS} &&&1.61$\times10^{-17} $&&5.40$\times10^{2} $  &\cite{Mousavipour2003}  \\     
63   &  \ce{H2S} &+& \ce{HO2} &$\rightarrow$& \ce{H2O} &+& \ce{HSO} &&&5$\times10^{-18} $&&  &\cite{Bulatov1990}  \\     
64   &  \ce{H2S} &+& \ce{CH3} &$\rightarrow$& \ce{CH4} &+& \ce{HS} &&&1.05$\times10^{-19} $&1.20&7.22$\times10^{2} $  &\cite{Mousavipour2003}  \\     
65   &  \ce{H2S} &+& \ce{SO2} &$\rightarrow$& \ce{H2O} &+& \ce{S2O} &&&1.09$\times10^{-19} $&1.86&1.90$\times10^{4} $  &\cite{Sendt2005}  \\     
66   &  \ce{H2S} &+& \ce{S2O} &$\rightarrow$& \ce{H2O} &+& \ce{S3} &&&7.08$\times10^{-19} $&1.51&1.71$\times10^{4} $  &\cite{Sendt2005}  \\    
67   &  \ce{SO} &+& \ce{HO2} &$\rightarrow$& \ce{SO2} &+& \ce{HO} &&&2.8$\times10^{-17} $&&  & \cite{Zhang2012}  \\        
68   &  \ce{SO} &+& \ce{S3} &$\rightarrow$& \ce{S2O} &+& \ce{S2} &&&1$\times10^{-18} $&&  &\cite{Moses2002}  \\             
69   &  \ce{SO} &+& \ce{O3} &$\rightarrow$& \ce{SO2} &+& \ce{O2} &&&4.5$\times10^{-18} $&&1.17$\times10^{3} $  &\cite{Atkinson2004}  \\     
70   &  \ce{SO} &+& \ce{O3} &$\rightarrow$& \ce{SO2} &+& \ce{O2(a^1\Delta g)} &&&3.6$\times10^{-19} $&&1.10$\times10^{3} $  &\cite{Sanders2006}  \\     
71   &  \ce{SO} &+& \ce{O2} &$\rightarrow$& \ce{SO2} &+& \ce{O} &&&4.37$\times10^{-20} $&1.40&1.87$\times10^{3} $  &\cite{Garland1998}  \\ 
72   &  \ce{SO} &+& \ce{CO2} &$\rightarrow$& \ce{SO2} &+& \ce{CO} &&&1.5$\times10^{-17} $&&2.20$\times10^{4} $ & \cite{Bauer1971}      \\          
73   &  \ce{SO} &+& \ce{NO2} &$\rightarrow$& \ce{SO2} &+& \ce{NO} &&&1.4$\times10^{-17} $&&  &\cite{Atkinson2004}  \\       
74   &  \ce{SO} &+& \ce{SO3} &$\rightarrow$& \ce{SO2} &+& \ce{SO2} &&&2$\times10^{-21} $&&  &\cite{Chung1975} \\    
75   &  \ce{SO2} &+& \ce{H} &$\rightarrow$& \ce{HO} &+& \ce{SO} &&&4.58$\times10^{-14} $&-2.30&1.56$\times10^{4} $  &\cite{Blitz2006}  \\     
76   &  \ce{SO2} &+& \ce{NO3} &$\rightarrow$& \ce{SO3} &+& \ce{NO2} &&&1.8$\times10^{-28} $&&  &\cite{Kurten2010}  \\     
77   &  \ce{SO2} &+& \ce{O3} &$\rightarrow$& \ce{SO3} &+& \ce{O2} &&&3.01$\times10^{-18} $&&7.00$\times10^{3} $  &\cite{Demore1997}  \\     
78   &  \ce{SO2} &+& \ce{HO2} &$\rightarrow$& \ce{HO} &+& \ce{SO3} &&&2.26$\times10^{-19} $&&3.42$\times10^{3} $  &\cite{Hwang2010}  \\     
79   &  \ce{SO2} &+& \ce{HO2} &$\rightarrow$& \ce{O2} &+& \ce{HOSO} &&&8.6$\times10^{-16} $&&5.23$\times10^{3} $  &\cite{Wang2005}  \\     
80   &  \ce{SO2} &+& \ce{O3} &$\rightarrow$& \ce{SO3} &+& \ce{O2(a^1\Delta g)} &&&6$\times10^{-20} $&&7.00$\times10^{3} $  &\cite{Zhang2012}  \\     
81   &  \ce{SO2} &+& \ce{O(^1D)} &$\rightarrow$& \ce{SO} &+& \ce{O2} &&&1.3$\times10^{-16} $&&  &\cite{Moses2002}  \\    
82   &  \ce{SO2} &+& \ce{NO2} &$\rightarrow$& \ce{SO3} &+& \ce{NO} &&&2$\times10^{-32} $&&  &\cite{Penzhorn1983}  \\     
83   &  \ce{SO3} &+& \ce{O} &$\rightarrow$& \ce{SO2} &+& \ce{O2} &&&1.06$\times10^{-19} $&2.57&1.47$\times10^{4} $  &\cite{Hindiyarti2007}  \\     
84   &  \ce{SO3} &+& \ce{H} &$\rightarrow$& \ce{HO} &+& \ce{SO2} &&&1.46$\times10^{-17} $&1.22&1.67$\times10^{3} $  &\cite{Hindiyarti2007}  \\     
85   &  \ce{SO3} &+& \ce{S2} &$\rightarrow$& \ce{S2O} &+& \ce{SO2} &&&2$\times10^{-22} $&&  &\cite{Moses2002}  \\     
86   &  \ce{SO3} &+& \ce{CO} &$\rightarrow$& \ce{CO2} &+& \ce{SO2} &&&1$\times10^{-17} $&&1.30$\times10^{4} $  &\cite{Krasnopolsky1994}   \\     
87   &  \ce{S2O} &+& \ce{O} &$\rightarrow$& \ce{SO} &+& \ce{SO} &&&1.7$\times10^{-18} $&&  &\cite{Mills1998}  \\            
88   &  \ce{S2O} &+& \ce{S2O} &$\rightarrow$& \ce{S3} &+& \ce{SO2} &&&1$\times10^{-20} $&&  &\cite{Mills1998}   \\      
89   &  \ce{OCS} &+& \ce{O} &$\rightarrow$& \ce{CO} &+& \ce{SO} &&&1.99$\times10^{-17} $&&2.15$\times10^{3} $  &\cite{Wei1975}       \\     
90   &  \ce{OCS} &+& \ce{O} &$\rightarrow$& \ce{CO2} &+& \ce{S} &&&8.3$\times10^{-17} $&&5.53$\times10^{3} $  &\cite{Singleton1988}       \\     
91   &  \ce{OCS} &+& \ce{C} &$\rightarrow$& \ce{CO} &+& \ce{CS} &&&1.01$\times10^{-16} $&&  &\cite{Dorthe1991}   \\       
92   &  \ce{OCS} &+& \ce{NO3} &$\rightarrow$& \ce{CO} &+& \ce{SO} &+& \ce{NO2} &1$\times10^{-22} $&&   &\cite{Atkinson2004}  \\     
93   &  \ce{OCS} &+& \ce{HO} &$\rightarrow$& \ce{CO2} &+& \ce{HS} &&&1.1$\times10^{-19} $&&1.20$\times10^{3} $  &\cite{Sanders2006}       \\     
94   &  \ce{CS} &+& \ce{O} &$\rightarrow$& \ce{CO} &+& \ce{S} &&&2.61$\times10^{-16} $&&7.58$\times10^{2} $  &\cite{Lilenfeld1977}   \\     
95   &  \ce{CS} &+& \ce{HO} &$\rightarrow$& \ce{H} &+& \ce{OCS} &&&1.7$\times10^{-16} $&&  &\cite{Wakelam2015}  \\     
96   &  \ce{CS} &+& \ce{HO} &$\rightarrow$& \ce{CO} &+& \ce{HS} &&&3$\times10^{-17} $&&  &\cite{Wakelam2015}  \\     
97   &  \ce{CS} &+& \ce{C} &$\rightarrow$& \ce{S} &+& \ce{C2} &&&1.44$\times10^{-17} $&0.5&2.04$\times10^{4} $  &\cite{Wakelam2015}  \\     
98   &  \ce{CS} &+& \ce{C2H3} &$\rightarrow$& \ce{H2C3S} &+& \ce{H} &&&1.7$\times10^{-18} $&&4.00$\times10^{2} $  &\cite{Wakelam2015}   \\     
99   &  \ce{CS} &+& \ce{CH} &$\rightarrow$& \ce{S} &+& \ce{C2H} &&&5$\times10^{-17} $&&  &\cite{Wakelam2015}  \\     
100   &  \ce{CS} &+& \ce{HN} &$\rightarrow$& \ce{S} &+& \ce{HNC} &&&1$\times10^{-17} $&&1.20$\times10^{3} $  &\cite{Wakelam2015}  \\     
101   &  \ce{CS} &+& \ce{NO2} &$\rightarrow$& \ce{OCS} &+& \ce{NO} &&&7.61$\times10^{-23} $&&  &\cite{Black1983}  \\     
102   &  \ce{CS} &+& \ce{O3} &$\rightarrow$& \ce{OCS} &+& \ce{O2} &&&3.01$\times10^{-22} $&&  &\cite{Black1983}  \\     
103   &  \ce{CS} &+& \ce{O2} &$\rightarrow$& \ce{OCS} &+& \ce{O} &&&2.62$\times10^{-22} $&&1.86$\times10^{3} $  &\cite{Richardson1975}  \\     
104   &  \ce{CS2} &+& \ce{O} &$\rightarrow$& \ce{CS} &+& \ce{SO} &&&2.76$\times10^{-17} $&&6.44$\times10^{2} $  &\cite{Wei1975}      \\     
105   &  \ce{CS2} &+& \ce{O} &$\rightarrow$& \ce{CO} &+& \ce{S2} &&&1.08$\times10^{-19} $&&  &\cite{Cooper1992}  \\     
106   &  \ce{CS2} &+& \ce{O} &$\rightarrow$& \ce{OCS} &+& \ce{S} &&&3.65$\times10^{-18} $&&5.83$\times10^{3} $  &\cite{Singleton1988}      \\     
107   &  \ce{CS2} &+& \ce{HO} &$\rightarrow$& \ce{OCS} &+& \ce{HS} &&&1.7$\times10^{-21} $&&  &\cite{Atkinson2004}  \\         
108   &  \ce{HSO3} &+& \ce{O2} &$\rightarrow$& \ce{HO2} &+& \ce{SO3} &&&1.3$\times10^{-18} $&&3.30$\times10^{2} $  &\cite{Atkinson1992}  \\     
109   &  \ce{HSO} &+& \ce{NO2} &$\rightarrow$& \ce{NO} &+& \ce{HSO2} &&&9.6$\times10^{-18} $&&  &\cite{Demore1997}  \\     
110   &  \ce{HSO} &+& \ce{O3} &$\rightarrow$& \ce{O2} &+& \ce{O2} &+& \ce{HS} &2.54$\times10^{-19} $&&3.84$\times10^{2} $ &\cite{Wang1990}   \\        
111   &  \ce{HSO2} &+& \ce{O2} &$\rightarrow$& \ce{HO2} &+& \ce{SO2} &&&3.01$\times10^{-19} $&&  &\cite{Demore1997}   \\     
112   &  \ce{HSOO} &+& \ce{O2} &$\rightarrow$& \ce{HO2} &+& \ce{SO2} &&&3.01$\times10^{-19} $&&  &\cite{Demore1997}   \\     
113   &  \ce{H2CS} &+& \ce{H} &$\rightarrow$& \ce{H2} &+& \ce{HCS} &&&9.33$\times10^{-17} $&&3.57$\times10^{3} $  &\cite{Vandeputte2010}  \\     
114   &  \ce{H2CS} &+& \ce{CH3} &$\rightarrow$& \ce{CH4} &+& \ce{HCS} &&&2.57$\times10^{-17} $&&4.93$\times10^{3} $  &\cite{Vandeputte2010}  \\      
115   &  \ce{CH3SH} &+& \ce{H} &$\rightarrow$& \ce{CH3} &+& \ce{H2S} &&&1.15$\times10^{-17} $&&8.41$\times10^{2} $  &\cite{Amano1983}   \\      
   \hline
\end{tabular}
\end{table*}

%%%%%%%%%%%%%%%%%%%%%%%%%%%%%%%%%%%%%%%%%%%%%%%%%%

%%%%%%%%%%%%%%%%% APPENDICES %%%%%%%%%%%%%%%%%%%%%

%\appendix


%%%%%%%%%%%%%%%%%%%%%%%%%%%%%%%%%%%%%%%%%%%%%%%%%%

% Don't change these lines
\bsp	% typesetting comment
\label{lastpage}
\end{document}